%% file: case4shifting-Arxiv.tex
\pdfoutput=1
\documentclass[preprints,article,accept,moreauthors,pdftex,10pt,a4paper]{mdpi}%preprints for Arxiv
\firstpage{1} 
\pubvolume{xx}
\issuenum{1}
\articlenumber{1}
\pubyear{2018}
\copyrightyear{2018}
\externaleditor{Academic Editor: name}
\history{Received: date; Accepted: date; Published: date}

\PassOptionsToPackage{inline}{enumitem}

\usepackage{amsfonts}
\usepackage{amssymb} %Symbols for maths.
\usepackage{mathtools}
\renewcommand{\intertext}[1]{\shortintertext{#1}}
\RequirePackage{etoolbox}
\providetoggle{extendedVersion}
\toggletrue{extendedVersion}
\togglefalse{extendedVersion}
\providetoggle{ieeeVersion}
\toggletrue{ieeeVersion}
\providetoggle{graphicalSummary}
\toggletrue{graphicalSummary}
\graphicspath{{../figs/}{../pics/}}
\usepackage{subcaption}

\usepackage{url}
\usepackage{pgfplots} %Ternary axis?
\pgfplotsset{compat=1.14} 
\usepgflibrary{shapes.misc} %eg. the clouds
\RequirePackage{tikz}%%This needs to be loaded after calling graphicx with
\usetikzlibrary{arrows, shapes,external, decorations.pathmorphing,
    decorations.pathreplacing, backgrounds, positioning, fit, pgfplots.ternary, pgfplots.units}
\usepackage{tabularx} %for the package on quantities around the Renyi entropy.
\usepackage{multirow} %to create different kinds of multiple rows.

\usepackage{xargs}                      % Use more than one optional parameter in a new commands
\usepackage[colorinlistoftodos,prependcaption,textsize=tiny]{todonotes}
\newcommandx{\unsure}[2][1=]{\todo[linecolor=red,backgroundcolor=red!25,bordercolor=red,#1]{#2}}
\newcommandx{\change}[2][1=]{\todo[linecolor=blue,backgroundcolor=blue!25,bordercolor=blue,#1]{#2}}
\newcommandx{\info}[2][1=]{\todo[linecolor=OliveGreen,backgroundcolor=OliveGreen!25,bordercolor=OliveGreen,#1]{#2}}
\newcommandx{\improvement}[2][1=]{\todo[linecolor=Plum,backgroundcolor=Plum!25,bordercolor=Plum,#1]{#2}}
\newcommandx{\thiswillnotshow}[2][1=]{\todo[disable,#1]{#2}}

\input{maxplus_minplus.tex}
\input{spectral_defs.tex}
\input{RenyiDefs.tex}

\Title{The case for shifting the Renyi Entropy}
\Author{Francisco J.~Valverde-Albacete$^{1,\ddagger}$\orcidA{}, Carmen~Pel\'aez-Moreno$^{2,\ddagger}$*\orcidB{}}

\AuthorNames{Francisco J.~Valverde-Albacete and Carmen~Pel\'aez-Moreno}

\address{%
$^{1}$ \quad Department of Signal Theory and Communications, Universidad Carlos III de Madrid, Legan\'es 28911, Spain; fva@tsc.uc3m.es\\
$^{2}$ \quad Department of Signal Theory and Communications, Universidad Carlos III de Madrid, Legan\'es 28911, Spain; carmen@tsc.uc3m.es}

\corres{Correspondence: carmen@tsc.uc3m.es; Tel.: +34-91-624-8771}

\firstnote{These authors contributed equally to this work.}
\abstract{%
We introduce a variant of the R\'enyi entropy definition that aligns it with the well-known H\"older mean: in the new formulation, the $r$-th order R\'enyi Entropy is the logarithm of the inverse of the $r$-th order H\"older mean. 
This brings about new insights into the relationship of the R\'enyi entropy to quantities close to it, like the information potential and the partition function of statistical mechanics. 
We also provide expressions that allow us to calculate the R\'enyi entropies from the Shannon cross-entropy and the escort probabilities. 
Finally, we discuss why shifting the R\'enyi entropy  is fruitful in some applications. 
}


\begin{document}
%%%%%%%%%%%%%%%%%%%%%%%%%%%%%%%%%%%%%%%%%%
%% Only for the journal Gels: Please place the Experimental Section after the Conclusions

\section{Introduction}
\label{sec:intro}
\input{intro.tex}

%\section{Materials and Methods}
\section{Preliminaries}
\label{sec:methods}

%\subsection{Means and averages}
\subsection{The Generalized Power Means}
\label{sec:means}
\input{means.tex}

\subsection{Renyi's entropy}
\label{sec:renyi:entropy}
\input{basicRenyi.tex}

\section{Results}
%\label{sec:results}
%\subsection{Shifting the R\'enyi entropy}
\label{sec:shifting}

\subsection{The shifted R\'enyi entropy and divergence}
%\subsection{Motivating the Renyi basic Information function}
\label{sec:rentropy:shifted}
\input{shiftedRenyiSemifields.tex}

%\todo[inline]{So far 3/11/18}
%\todo[inline]{So far 9/11/18}

\subsection{Writing R\'enyi entropies in terms of each other}
\label{sec:entropy:rewriting}
\input{rewritingRE.tex}

\subsection{Quantities around the shifted R\'enyi entropy}
\label{sec:ca:rentropy}
\input{circaRentropy.tex}

\subsection{Discussion}
\label{sec:discussion}
\input{discussionC4Shifting.tex}

\section{Conclusions}
\label{sec:conc}
\input{conclusionsC4S.tex}
\vspace{6pt} 

%%%%%%%%%%%%%%%%%%%%%%%%%%%%%%%%%%%%%%%%%%
%% optional
%\supplementary{The following are available online at \linksupplementary{s1}, Figure S1: title, Table S1: title, Video S1: title.}

% \supplementary{The following are available at \linksupplementary, Figure S1: title, Table S1: title, Video S1: title. A supporting video article is available at doi: link.}

%%%%%%%%%%%%%%%%%%%%%%%%%%%%%%%%%%%%%%%%%%
%\authorcontributions{For research articles with several authors, a short paragraph specifying their individual contributions must be provided. The following statements should be used ``Conceptualization, X.X. and Y.Y.; Methodology, X.X.; Software, X.X.; Validation, X.X., Y.Y. and Z.Z.; Formal Analysis, X.X.; Investigation, X.X.; Resources, X.X.; Data Curation, X.X.; Writing—Original Draft Preparation, X.X.; Writing—Review \& Editing, X.X.; Visualization, X.X.; Supervision, X.X.; Project Administration, X.X.; Funding Acquisition, Y.Y.'', please turn to the \href{http://img.mdpi.org/data/contributor-role-instruction.pdf}{CRediT taxonomy} for the term explanation. Authorship must be limited to those who have contributed substantially to the work reported. }
%\authorcontributions{Software: F.V.A.; all other contributions: F.V.A. and C.P.M.}

\authorcontributions{Conceptualization, Francisco J Valverde-Albacete and Carmen Pel\'aez-Moreno; Formal analysis, Francisco J Valverde-Albacete and Carmen Pel\'aez-Moreno; Funding acquisition, Carmen Pel\'aez-Moreno; Investigation, Francisco J Valverde-Albacete and Carmen Pel\'aez-Moreno; Methodology, Francisco J Valverde-Albacete and Carmen Pel\'aez-Moreno; Software, Francisco J Valverde-Albacete; Supervision, Carmen Pel\'aez-Moreno; Validation, Francisco J Valverde-Albacete and Carmen Pel\'aez-Moreno; Visualization, Francisco J Valverde-Albacete and Carmen Pel\'aez-Moreno; Writing – original draft, Francisco J Valverde-Albacete and Carmen Pel\'aez-Moreno; Writing – review \& editing, Francisco J Valverde-Albacete and Carmen Pel\'aez-Moreno.}

%%%%%%%%%%%%%%%%%%%%%%%%%%%%%%%%%%%%%%%%%%
%\funding{Please add: ``This research received no external funding'' or ``This research was funded by [name of funder] grant number [xxx].'' Check carefully that the details given are accurate and use the standard spelling of funding agency names at \url{https://search.crossref.org/funding}, any errors may affect your future funding.}
\funding{This research was funded by he Spanish Government-MinECo projects
TEC2014-53390-P %Excelencia y retos, antiguos CICYT
and 
TEC2017-84395-P %Excelencia y retos, antiguos CICYT
%TEC2014-61729-EXP.%El explora.
}

%%%%%%%%%%%%%%%%%%%%%%%%%%%%%%%%%%%%%%%%%%
%\acknowledgments{In this section you can acknowledge any support given which is not covered by the author contribution or funding sections. This may include administrative and technical support, or donations in kind (e.g. materials used for experiments).}

%%%%%%%%%%%%%%%%%%%%%%%%%%%%%%%%%%%%%%%%%%
%\conflictsofinterest{Declare conflicts of interest or state ``The authors declare no conflict of interest.'' Authors must identify and declare any personal circumstances or interest that may be perceived as inappropriately influencing the representation or interpretation of reported research results. Any role of the funding sponsors in the design of the study; in the collection, analyses or interpretation of data; in the writing of the manuscript, or in the decision to publish the results must be declared in this section. If there is no role, please state ``The founding sponsors had no role in the design of the study; in the collection, analyses, or interpretation of data; in the writing of the manuscript, and in the decision to publish the results''.} 
\conflictofinterest{The authors declare no conflict of interest.}

%%%%%%%%%%%%%%%%%%%%%%%%%%%%%%%%%%%%%%%%%%
%% optional
%\abbreviations{The following abbreviations are used in this manuscript:\\
%
%\noindent 
%\begin{tabular}{@{}ll}
%PCA & Principal Component Analysis\\
%ICA & Independent Component Analysis\\
%CMET & Channel Multivariate Entropy Triangle\\
%CBET & Channel Binary Entropy Triangle\\
%SMET & Source Multivariate Entropy Triangle
%%MDPI & Multidisciplinary Digital Publishing Institute\\
%%DOAJ & Directory of open access journals\\
%%TLA & Three letter acronym\\
%%LD & linear dichroism
%\end{tabular}}

%%%%%%%%%%%%%%%%%%%%%%%%%%%%%%%%%%%%%%%%%%%
%%% optional
%\appendixtitles{no} %Leave argument "no" if all appendix headings stay EMPTY (then no dot is printed after "Appendix A"). If the appendix sections contain a heading then change the argument to "yes".
\appendixsections{multiple} %Leave argument "multiple" if there are multiple sections. Then a counter is printed ("Appendix A"). If there is only one appendix section then change the argument to "one" and no counter is printed ("Appendix").
\appendix
\section{The Kolmogorov-mean and the Kolmogorov-Nagumo formula}
\label{sec:KN}
\input{KNmean.tex}

\section{The approach to Shannon's information functions based in postulates}
\label{sec:Shannon}
\input{postulatesShannon.tex}

\externalbibliography{yes}
%\bibliography{your_external_BibTeX_file}
\bibliography{case4shifting}
\end{document}

%% file: spectral_defs.tex
\newcommand{\supp}{\ensuremath\operatorname{supp}}

\RequirePackage{bbold}%blackboard bold for the lattice names

%% file: RenyiDefs.tex
\newcommand{\hmean}[3][r]{\ensuremath{M_{#1}(#2,#3)}}

\newcommand{\pentropy}[2][r]{\ensuremath{\tilde H_{#1}(#2)}}
\newcommand{\pdiv}[3][r]{\ensuremath{\tilde D_{#1}(#2\|#3)}}

\newcommand{\hinf}[1]{\ensuremath{\mathfrak I_\ast(#1) }}
\newcommand{\hinfinv}[1]{\ensuremath{\left (\mathfrak I_\ast\right)^{-1}(#1)}}
\newcommand{\sentropy}[1]{\ensuremath{H(#1)}} % Shannon entropy
\newcommand{\sdiv}[2]{\ensuremath{D_{KL}{(#1\|#2)}}} %KL-divergence
\newcommand{\rentropy}[2][\alpha]{\ensuremath{H_{#1}(#2)}}
\newcommand{\rdiv}[3][\alpha]{\ensuremath{D_{#1}(#2\|#3)}}

\newcommand{\rdist}[2][\alpha]{\ensuremath{q_{#1}(#2)}} %rescort!

\newcommand{\srprob}[2][r]{\ensuremath{\tilde P_{#1} \left (#2\right)}}
\newcommand{\srpot}[2][r]{\ensuremath{\tilde V_{#1} \left (#2\right)}}
\newcommand{\srentropy}[2][r]{\ensuremath{\tilde H_{#1} \left (#2\right)}}
\newcommand{\srcrossent}[3][r]{\ensuremath{\tilde X_{#1} \left (#2\|#3\right)}}
\newcommand{\srdiv}[3][r]{\ensuremath{\tilde D_{#1}\left(#2\|#3 \right)}}
\newcommand{\srdist}[2][r]{\ensuremath{\tilde q_{#1}(#2)}} %shifted rescort!

%% file: intro.tex
Let $X\sim P_X$ be a random variable over a set of outcomes $\mathcal X = \{x_i\mid 1\leq i \leq n\}$ and \emph{pmf} $P_X$ defined  in terms of the non-null values $p_i = P_X(x_i)$\,. 
The R\'enyi entropy for $X$  is defined in terms of that of $P_X$ as $\rentropy{X} = \rentropy{P_X}$ by a case analysis~\cite{ren:61}
\begin{align}
%\label{eq:RenyiE:intro}
\label{eq:RenyiE:def}
\alpha &\neq 1 
\quad
 \rentropy{P_X} = \frac{1}{1-\alpha}\log\sum_{i=1}^n {p_i^\alpha} \quad \\
\alpha &= 1 
\quad
\lim_{\alpha \rightarrow 1}  \rentropy{P_X} = \sentropy{P_X}
\notag
%= - \sum_i p_i \log {p_i} 	\notag
%
\end{align}
where $\sentropy{P_X} =  - \sum_{i=1}^n  {p_i} \log p_i$ is the Shannon entropy~\cite{sha:wea:49,sha:48a,sha:48b}. 
Similarly the associated divergence when $Q\sim Q_X$ is substituted by $P\sim P_X$ on a compatible support  is defined in terms of their \emph{pmf}s  $q_i = Q_X(x_i)$ and $p_i = P_X(x_i)$, respectively, as $\rdiv{X}{Q}  = \rdiv{P_X}{Q_X} $ where 
\begin{align}
\label{eq:RenyD:def} 
\alpha &\neq 1 
\quad 
\rdiv{P_X}{Q_X} = \frac{1}{\alpha-1}\log\sum_{i=1}^n p_i^\alpha q_i^{1- \alpha}\\
\alpha &= 1 
\quad 
 \lim_{\alpha \rightarrow 1}  \rdiv{P_X}{Q_X} = \sdiv{P_X}{Q_X}.  \notag	
%	= \frac{1}{\alpha-1}\log\sum_{i=1}^np_i \left(\frac{p_i}{q_i}\right)^{\alpha -1}
%
\end{align}
%$\sentropy{P_X}= \lim_{\alpha \rightarrow 1}  \rentropy{P_X}$  % are the Shannon entropy and 
and 
$\sdiv{P_X}{Q_X}= \sum_{i=1}^n  {p_i} \log \frac{p_i}{q_i}$
is the Kullback-Leibler divergence~\cite{erv:har:14}.

When trying to find the closed form for a generalization of the Shannon entropy that was compatible with all the Faddev axioms but that of linear average, R\'enyi found that the function $\varphi(x)=x^r$ could be used with the Kolmogorov-Nagumo average to obtain such a new form of entropy. 
Rather arbitrarily, he decided that the constant should be $\alpha = r + 1$\,, thus obtaining \eqref{eq:RenyiE:def} and \eqref{eq:RenyD:def}, but obscuring the relationship of the entropies of order $\alpha$ and the generalized power means. 

We propose to shift the parameter in these definitions back to $r=\alpha -1$ to define %so that %the new expression read:
%Let $X$ be a discrete random variable with \emph{pmf} $P_X$. 
%We call
%\begin{itemize}
%\item 
the \emph{shifted R\'enyi entropy of order $r$} the value
\begin{align}
\label{eq:sRenyE}
\pentropy{P_X} &= - \log \hmean{P_X}{P_X}
%\end{align}
\intertext{and the \emph{shifted R\'enyi divergence of order $r$} the value}
%\item 
%\begin{align}
\label{eq:sRenyD}
\pdiv{P_X}{Q_X}\ &= \log \hmean{P_X}{\frac{P_X}{Q_X}}
\intertext{where $M_r$ is the \emph{$r$-th order weighted generalized power means} or \emph{H\"older means}~\cite{har:lit:pol:52}:
}
\label{eq:holder:wmean}
\hmean{\vec w}{\vec x} &= \left( \sum_{i=1}^n \frac{w_i}{\sum_k w_k}\cdot x_i^r\right)^\frac{1}{r}.
%&
%\lim_{r\rightarrow 0}\hmean{\vec w}{\vec x} &= %\hmean[0]{\vec w}{\vec x} = 
%\left(\prod_{i=1}^{n} x_i^{w_i}\right)^\frac{1}{\sum_k w_k}.
\end{align}
%\end{itemize}

Since this could be deemed equally arbitrary, in this paper we argue
that this statement of the R\'enyi entropy greatly clarifies its role vis-a-vis the H\"older means,  
%and the Minkowsky $p$-norm, $\| \vec x \|_p=(\sum_{i=1}^n x_i^p)^{1/p}$, 
viz. that most of the properties and special cases of the R\'enyi entropy arise from similar concerns in the H\"older means. 
We also provide a brief picture of how would the theory surrounding the R\'enyi entropy be modified with this change, as well as its relationship to some other magnitudes. 

%For this purpose, we first revisit the generalized power means and take a detailed look at their properties. 
%We next revisit the core of the theory of the R\'enyi entropy over discrete variables, and then restate all of these results in terms of the means.
%We conclude with a discussion of similar attempts to relate the R\'enyi entropy to different ``core'' functions and why we believe the H\"older means are a superior starting point.

%% file: means.tex
%\subsubsection{The weighted power means}
%\label{sec:holder_mean}
%
Recall that the \emph{generalized power or H\"older mean of order $r$} is defined as\footnote{In this paper we use the notation where the weighting vector comes first---rather than the opposite, used in~\cite{har:lit:pol:52}---to align it with formulas in information theory, e.g. divergences and cross entropies.} 
\begin{align}
\label{eq:holder:wmean}
\hmean{\vec w}{\vec x} = \left( \frac{\sum_{i=1}^n w_i\cdot x_i^r}{\sum_k w_k}\right)^\frac{1}{r}
= \left( \sum_{i=1}^n \frac{w_i}{\sum_k w_k}\cdot x_i^r\right)^\frac{1}{r}
\end{align}
%Due to the homogeneity about to be discovered below, without loss of generality we suppose that $\sum w_i =1$ and we will identify these weights with a probability mass function $p=[p_i]_{i=1}^n$\,. 
By formal identification, 
%Note that, in such case,  
the generalized power mean is nothing but the weighted $f$-mean with $f(x) = x^r$ %,  and $w_i = \frac{1}{n}$ 
(see Appendix~\ref{sec:KN}). %we get the Kolmogorov mean. 
%The author of 
Reference~\cite{kit:34} provides proof that this functional mean also has the properties \ref{prop:knmean:9}-\ref{prop:knmean:8} of Proposition~\ref{propo:KN} and Associativity. 

%\begin{example}
%Due to property~\ref{propo:prop:hmean}.\ref{prop:hmean:5} , the 
The 
evolution of $\hmean{\vec w}{\vec x}$ with $r $ is also called the \emph{H\"older path (of an $\vec x$)}. 
Important cases of this mean for historical and practical reasons are obtained by giving values to $r$:
\begin{itemize}
\item The (weighted) geometric mean when $r=0$.
%\[
%\hmean[0]{\vec w}{\vec x} = \lim_{r\rightarrow 0} \hmean{\vec w}{\vec x} = \left(\Pi_{i=1}^n x_i^{w_i}\right)^\frac{1}{\sum_k w_k}
%\]
\begin{align}
\label{eq:geom:mean}
\hmean[0]{\vec w}{\vec x} = \lim_{r\rightarrow 0} \hmean{\vec w}{\vec x} = \left(\Pi_{i=1}^n x_i^{w_i}\right)^\frac{1}{\sum_k w_k}
\end{align}
\item The weighted arithmetic mean when $r=1$.
\[
\hmean[1]{\vec w}{\vec x} = \sum_{i=1}^n \frac{w_i}{\sum_k w_k}\cdot x_i
\]

\item The weighted harmonic mean for $r=-1$. 
\[
\hmean[-1]{\vec w}{\vec x} = \left( \sum_{i=1}^n \frac{w_i}{\sum_k w_k}\cdot x_i^{-1}\right)^{-1}
			= \frac{\sum_k w_k}{\sum_{i=1}^n w_i\cdot \frac{1}{x_i}}
\]
\item  The quadratic mean for $r=2$.
\[
\hmean[2]{\vec w}{\vec x} = \left( \sum_{i=1}^n \frac{w_i}{\sum_k w_k}\cdot x_i^2\right)^\frac{1}{2}
\]
\item Finally, the max- and min-means appear as the limits:
\begin{align*}
\hmean[\infty]{\vec w}{\vec x} &= \lim_{r\rightarrow \infty} \hmean{\vec w}{\vec x} = \max_{i=1}^n x_i
\\
\hmean[-\infty]{\vec w}{\vec x} &= \lim_{r\rightarrow -\infty} \hmean{\vec w}{\vec x} = \min_{i=1}^n x_i
\end{align*}
\end{itemize}

%\todo[inline]{Prove what axioms of the KN means are valid for weighted means.!}

% % % % % % % % % % % % % % % % % % % % % % % % % % % % % % % % % % % % % % % % % % %
% % % % PROPERTIES OF THE POWER MEANS
% % % % % % % % % % % % % % % % % % % % % % % % % % % % % % % % % % % % % % % % % % %
%Furthermore, the weighted means 
They all show the following properties:
\begin{Proposition}[Properties of the weighted power means]
\label{propo:prop:hmean}
Let $\vec x, \vec w \in (0,\infty)^n$ and $r,s \in (-\infty,\infty)$\,.
%$r,s \in (-\infty,0) \cup (0,\infty)$\,. 
%Let $\vec w, \vec x \in {(\nnR})^n$ and $r,s \in \mathbb R\setminus0$\,. 
 Then, the following formal identities hold,
 where %$\hmean[-1]{\cdot}{\cdot}$ is the weighted harmonic mean and 
 ${\vec x}^r$ and $\frac{1}{\vec x}$ are to be understood entry-wise, 
\begin{enumerate}
\item \label{prop:hmean:1} ({0- and 1-order homogeneity in weights and values}) If $k_1, k_2 \in \nnR$\,, then 
$
\hmean{k_1\cdot \vec w}{k_2\cdot \vec x} = k_1^0 \cdot k_2^1\cdot \hmean{\vec w}{\vec x}\,.
$

\item \label{prop:hmean:2}  ({Order factorization})
If $r \neq 0 \neq s$, then 
$\hmean[rs]{\vec w}{\vec x} = \left( \hmean[s]{\vec w}{(\vec x)^r}\right)^{1/r}$

\item \label{prop:hmean:3} ({Reduction to the arithmetic mean})
If $r \neq 0$, then 
$
\hmean[r]{\vec w}{\vec x} = [\hmean[1]{\vec w}{(\vec x)^r}]^{1/r}
$\,.
%where $\hmean[1]{\cdot}{\cdot}$ is the weighted arithmetic mean. 

\item \label{prop:hmean:4} ({Reduction to the harmonic mean})
If $r \neq 0$, then 
$
\hmean[-r]{\vec w}{\vec x} = [\hmean[-1]{\vec w}{(\vec x)^r}]^{1/r}
	= [\hmean[r]{\vec w}{\frac{1}{\vec x}}]^{-1} 
	= [\hmean[1]{\vec w}{\frac{1}{(\vec x)^r}}]^{-1/r}
$\,.

\item \label{prop:hmean:5} 
({Monotonicity in $r$}) 
If $\vec x \in [0,\infty]^n$ %$\vec w, \vec x \in {(\nnR})^n$ and 
and $r,s \in [-\infty,\infty]$\,, then 
$$\min_i x_i = \hmean[-\infty]{\vec w}{\vec x} \leq \hmean{\vec w}{\vec x} \leq \hmean[\infty]{\vec w}{\vec x} = \max_i x_i$$ 
and the mean is a strictly monotonic function of $r$, that is 
$r < s$ implies 
 $
 \hmean[r]{\vec w}{\vec x} < \hmean[s]{\vec w}{\vec x}
 %M_p(\vec w, \vec x) \leq M_q (\vec w, \vec x)\,.
 $, 
 unless:
 \begin{itemize}
 \item $x_i = k$ is constant, in which case $\hmean[r]{\vec w}{\vec x} = \hmean[s]{\vec w}{\vec x} = k$\,.
 \item $s \leq 0$ and some $x_i = 0$, in which case $0 = \hmean[r]{\vec w}{\vec x} \leq \hmean[s]{\vec w}{\vec x}$\,.
 \item $0 \leq r$ and some $x_i = \infty$, in which case $\hmean[r]{\vec w}{\vec x} \leq \hmean[s]{\vec w}{\vec x} = \infty$\,.
 \end{itemize}
%\item \label{prop:hmean:2} The series of means are bounded by the extremes in the series:
%\[
%\min_i p_i\cdot x_i \leq \hmean{\vec w}{x} \leq \max_i p_i \cdot x_i
%\]

%\item \label{prop:hmean:6} \textbf{Monotonicity in $r$}. 
\item \label{prop:hmean:6} ({Non-null derivative})  
Call $\srdist[r]{\vec w, \vec x} %= \{w'_k\}_{k=1}^n 
= \left\{\frac{w_k x_k^r}{\sum_i w_i x_i^r}\right\}_{k=1}^n$\,. Then
\begin{align}
\label{eq:mean:derivative}
\frac{\delta}{\delta{r}}\hmean{\vec w}{\vec x} 
= \frac{1}{r} \cdot \hmean[r]{\vec w}{\vec x} \ln \frac{\hmean[0]{\srdist[r]{\vec w, \vec x}}{\vec x}}{\hmean[r]{\vec w}{\vec x}}
\end{align}
\end{enumerate}
\end{Proposition}
\begin{proof}
Property~\ref{prop:hmean:1}  follows from the commutativity, associativity and cancellation of sums and products in $\nnR$. 
Property~\ref{prop:hmean:2} follows from identification in the definition, then properties~\ref{prop:hmean:3} and~\ref{prop:hmean:4} follow from it with $s=1$ and $s=-1$ respectively. %\ref{prop:hmean:2}. 
Property~\ref{prop:hmean:5} and the special cases in it are well known and studied extensively  in~\cite{har:lit:pol:52}. 
We will next prove property~\ref{prop:hmean:6}
% % % EXAMPLE of how to split equations between lines from
% https://tex.stackexchange.com/questions/44450/how-to-align-a-set-of-multiline-equations
%\newcommand{\myvec}[1]{\hat{\mathbf{#1}}}% Vector notation
%\begin{align}
%  f_{\textit{P},\textit{P}}\left(\myvec{n};\myvec{m}\right) &= \frac{\omega^2}{4\pi\rho\alpha^4} \textit{AF}\left(k_\alpha\left(\myvec{n}-\myvec{m}\right)\right) \nonumber \\
%    &\mathrel{\phantom{=}} \times\left\{\left(\lambda+\mu\right)^2\eta_N+\left(\lambda+\mu\right)\mu\eta_N\left(\cos 2\phi+\cos 2\theta\right)\right. \nonumber \\
%    &\mathrel{\phantom{=}} \left.\kern-\nulldelimiterspace +\;\mu^2\eta_N\cos 2\phi\cos 2\theta+\mu^2\eta_T\sin 2\phi\sin 2\theta\cos\varphi\vphantom{\left(\lambda\right)^2}\right\}, \\
%  f_{\textit{P},\textit{SH}}\left(\myvec{n};\myvec{m},\myvec{q}\right) &= \frac{\omega^2}{4\pi\rho\alpha\beta^3} \textit{AF}\left(k_\alpha\myvec{n}-k_\beta\myvec{m}\right) \nonumber \\
%    &\mathrel{\phantom{=}} \times\left(-\mu^2\eta_T\right)\sin 2\phi\cos\theta\sin\varphi, \\
%  f_{\textit{P},\textit{SV}}\left(\myvec{n};\myvec{m},\myvec{q}\right) &= \frac{\omega^2}{4\pi\rho\alpha\beta^3} \textit{AF}\left(k_\alpha\myvec{n}-k_\beta\myvec{m}\right) \nonumber \\
%    &\mathrel{\phantom{=}} \times\left\{\left(\lambda+\mu\right)\mu\eta_N\sin 2\theta+\mu^2\eta_N\cos 2\phi\sin 2\theta\right. \nonumber \\
%    &\mathrel{\phantom{=}} \left.\kern-\nulldelimiterspace -\;\mu^2\eta_T\sin 2\phi\cos 2\theta\cos\varphi\right\},
%\end{align}
%\todo[inline]{Correct equations below with CPMs review copy.}
\begin{align*}
\frac{d}{dr}\hmean[r]{\vec w}{\vec x} 
&= \frac{d}{dr} e^{\frac{1}{r}\ln\left(\sum_k \frac{w_k}{\sum_i w_i} x_k^r\right )}  
= \hmean[r]{\vec w}{\vec x} %\cdot %\\
%&\mathrel{\phantom{=}}  \cdot 
\left( \frac{-1}{r^2}\ln\left(\sum_k \frac{w_k}{\sum_i w_i}  x_k^r\right ) + \frac{1}{r}\cdot\frac{\sum_k w_k x_k^r \ln x_k}{\sum_i w_i x_i^r}  \right)
\end{align*}
Note that if we call 
$\srdist[r]{\vec w, \vec x}  = \{w'_k\}_{k=1}^n = \left\{\frac{w_k x_k^r}{\sum_i w_i x_i^r}\right\}_{k=1}^n$\,, 
since this is a probability  we may rewrite:
\begin{align*}
\sum_k  \frac{w_k x_k^r }{\sum_i w_i x_i^r} \cdot \ln x_k
= \sum_k w'_k \ln x_k 
= \ln \left(\prod_k x_k^{w'_k}\right) 
= \ln \hmean[0]{\srdist[r]{\vec w, \vec x}}{x} 
\end{align*}
whence
\begin{align*}
\frac{d}{dr}\hmean[r]{\vec w}{\vec x} 
&= \hmean[r]{\vec w}{\vec x} \left( \frac{1}{r}\cdot \ln \hmean[0]{\srdist[r]{\vec w, \vec x}}{x} %+ \right. \\
				 %&\mathrel{\phantom{=}} \left.\kern-\nulldelimiterspace 
				 - \frac{1}{r}\cdot \ln \hmean[r]{\vec w}{\vec x}  \right) =
%\notag 
\\
&=  \frac{1}{r} \cdot \hmean[r]{\vec w}{\vec x} \ln \frac{\hmean[0]{\srdist[r]{\vec w, \vec x}}{\vec x}}{\hmean[r]{\vec w}{\vec x}}\,.
\end{align*}
\end{proof}
%And perhaps the best-known properties
%\begin{proposition}[Monotonicity of the means in $r$]
%Let $p, x \in {(\nnR})^n$ and $r,s \in [-\infty,\infty]$\,. 
%If $r < s$, then 
% $
% \hmean[r]{\vec w}{x} < \hmean[s]{\vec w}{x}
% %M_p(\vec w, \vec x) \leq M_q (\vec w, \vec x)\,.
% $
% unless:
% \begin{itemize}
% \item $x_i = k$ is constant, in which case $\hmean[r]{\vec w}{x} = \hmean[s]{\vec w}{x} = k$\,.
% \item $s \leq 0$ and some $x_i = 0$, in which case $0 = \hmean[r]{\vec w}{x} \leq \hmean[s]{\vec w}{x}$\,.
% \item $0 \leq r$ and some $x_i = \infty$, in which case $\hmean[r]{\vec w}{x} \leq \hmean[s]{\vec w}{x} = \infty$\,.
% \end{itemize}
% In particular, $\hmean[-\infty]{\vec w}{x} \leq \hmean{\vec w}{x} \leq \hmean[\infty]{\vec w}{x}$\,.
%\end{proposition}
\begin{Remark}
The distribution $\srdist[r]{\vec w, \vec x}$ when $\vec w = \vec x$ is extremely important in the theory of generalized entropy functions, where it is called a  \emph{(shifted) escort distribution (of $\vec w$)}~\cite{bec:sch:95}, and we will prove below that its importance stems, at leasts partially, from this property. 
\end{Remark}

\begin{Remark}
\label{rem:disc}
Notice that in the case where both conditions at the end of Property~\ref{propo:prop:hmean}.\ref{prop:hmean:5} hold---that is for $i \neq j$ we have $x_i = 0$ and $x_j=\infty$---then we have for $r \leq 0,\,\hmean[r]{\vec w}{\vec x} = 0$ and for $0 \leq r\,, \hmean[r]{\vec w}{\vec x} = \infty$ whence $\hmean[r]{\vec w}{\vec x}$ has a discontinuity at $r=0$.
\end{Remark}

%% file: basicRenyi.tex
Although the following material is fairly standard, it bears directly into our discussion, hence we introduce it in full.

\subsubsection{Probability spaces, random variables and expectations}
\label{sec:prob}
\input{probRRVV.tex}

\subsubsection{The approach to R\'enyi's information functions based in postulates}
%\subsubsection{Moments of random variables as means}
%%\todo{Relate to the \emph{R\'enyi spectrum of (x)}}.
%\todo[inline]{Remember to include expectations of a function of a R.V.}
%
One of the most  important applications of the generalized weighted means is to calculate the moments of (non-negative) random variables.
\begin{Lemma}
Let $X \sim P_X$ be a discrete random variable. Then the \emph{$r$-th moment of $X$} is:
\begin{align}
E_X\{ X^r\} = \sum_i p_i  x_i^r = \left(M_r(P_X, X)\right)^r
\end{align}
\end{Lemma}
\noindent
This is the concept that Shannon, and afterwards R\'enyi, used to quantify information by using the distribution as a random variable (\S~\ref{sec:ca:rentropy}). 

The postulate approach to characterize Shannon's information measures can be found in Appendix~\ref{sec:Shannon}. 
Analogue generalized postulates lead to R\'enyi's information functions, but, importantly, he did \emph{not} consider normalized measures, that is with $\sum_k p_k = 1$. 

We follow~\cite{jiz:ari:04} in stating the R\'enyi postulates:
\begin{enumerate}
\item \label{pos:Ren:1} The amount of information provided by a single random event $x_k$ should be a function of its probability $P_X(x_k) = p_k$, not its value $x_k=X(\omega_k)$, $\mathfrak I: [0,1] \rightarrow I $ where $I \subseteq \mathbb R$ quantifies information. 
%\begin{align}
%\mathfrak I(x_k) = \mathfrak I (p_k), p_k \in [0,1]
%\end{align}

\item \label{pos:Ren:2} This amount of information should be additive on independent events.
\begin{align}
\label{eq:Cauchy's}
\mathfrak I(p,q) = \mathfrak I(p) + \mathfrak I(q)
\end{align}

\item \label{pos:Ren:3}  %\textbf{Fixing the unit for the amount of information: the bit} 
The amount of information of a binary equiprobable decision is one bit.
\begin{align}
\mathfrak I(1/2) = 1
\end{align}

\item \label{pos:Ren:4} If different amounts of information occur with different probabilities the total amount of information $\mathfrak I$ is an \emph{average} of the individual information amounts weighted by the probability of occurrence.
\end{enumerate}
%Note how the last postulate describes aggregate amounts of information, not atomic ones. 

These postulates \emph{may} lead to the following consequences:
\begin{itemize}
\item Postulates~\ref{pos:Ren:1} and~\ref{pos:Ren:2} fix \emph{Hartley's function} as the single possible amount of information of a basic event 
\begin{align}
\label{eq:Hartley_s}
\mathfrak I: [0,1] \rightarrow [0,\infty], p \mapsto \mathfrak I(p) = -k \log p.
\end{align}

\item Postulates~\ref{pos:Ren:3} fixes the base of the logarithm in Hartley's formula to $2$ by fixing $k=1$. Any other value $k = 1/\log b$ fixes b as the base for the logarithm and changes the unit.  

\item Postulate~\ref{pos:Ren:4} defines an {average amount of information}, or \emph{entropy}, properly speaking%
\footnote{The ``entropy'' in Information Theory is, by definition, synonym with ``aggregate amount of information'', which departs from its physical etymology, despite the numerous analogies between both concepts.}.
Its basic formula is a form of the 
Kolmogorov-Nagumo formula or $f$-mean~\eqref{eq:KN:formula} 
applied to information
\begin{align}
\label{eq:KNentropy}
%H(P) = 
H(P_X, \varphi, \mathfrak I) = \varphi^{-1}\left(\ \Sigma_{i=1}^n \frac{p_i}{\sum_k p_k} \varphi(\mathfrak I(p_i))\right) \,.
\end{align}
%Notice that this is a way to define an average in this semifield to obtain the R\'enyi entropy. 
It has repeatedly been proven that only two forms of the function $\varphi$ can actually be used in the Kolmogorov-Nagumo formula that respect the previous postulates~\cite{ren:61,ren:70,jiz:ari:04}:
\begin{itemize}
\item The one generating Shannon's entropy: %, in R\'enyi's framework
\begin{align}
\label{eq:linear}
\varphi(h)=ah + b\,\text{with}\,a\neq 0\,,
\end{align}
\item  That originally used by R\'enyi himself:
\begin{align}
\label{eq:exponential}
\varphi(h) = 2^{(1 -\alpha)h}, \, \text{with}\, \alpha \in [-\infty, \infty]\setminus\{1\}\,.
\end{align} 
\end{itemize}
\end{itemize}

%These decisions lead to the following formulae for the generalized entropy for a random variable $X \sim  P_X$. %that we introduced without motivation in Section~\ref{sec:intro}, 
Taking the first form~\eqref{eq:linear} and plugging it into~\eqref{eq:KNentropy} leads to \emph{Shannon's measure of information}, 
%\begin{align*}
%%\pentropy[0]{p_x} 
%H(P_X) &= \frac{1}{a}\left(\sum_i \frac{p_i}{\sum_k p_k}(a \log \frac{1}{p_i} + b) - b\right)
%	=  \frac{1}{a}\left(\sum_i \frac{p_i}{\sum_k p_k} a \log \frac{1}{p_i}\right) %=  \sum_i p_i \log \frac{1}{p_i}\\
%	= - \sum_i \frac{p_i}{\sum_k p_k} \log {p_i}\,, %= - \log \prod_i p_i^{p_i}\,,
%%	\tag{\ref{eq:renyi:entropy:ori} revisited}
%	%= \log \frac{1}{\}
%\end{align*}
%to obtain Shannon's entropy, but we will see soon that this is not necessary.
and taking the second form leads to \emph{R\'enyi's measure of information} \eqref{eq:RenyiE:def}, so we actually have: 
\begin{Definition}[\cite{ren:61,ren:70}] %[R\'enyi entropy]
The \emph{R\'enyi entropy of order $\alpha$} for a discrete random variable $ X \sim P_X$, is 
\begin{align}
\label{eq:RenyE:unfolded}
&%H_\alpha(X) = 
H_\alpha(P_X) = \frac{1}{1-\alpha}\log\left(\sum_{i=1}^n \frac{p_i^\alpha}{\sum_k p_k}\right)\quad \alpha \neq 1
%\tag{\ref{eq:RenyE} revisited}
%\\
%
%\label{eq:shannon:as:renyi}
&%\lim_{\alpha \rightarrow 1} H_\alpha(X) = 
\lim_{\alpha \rightarrow 1}  H_\alpha(P_X)  
	= H(P_X) = - \sum_i \frac{p_i}{\sum_k p_k}\log p_i\,,
%\tag{\ref{eq:shannon:as:renyi} revisited}
%
\end{align}
where the fact that Shannon's entropy is the R\'enyi entropy when $\alpha \rightarrow 1$ in \eqref{eq:RenyiE:def} is found by a continuity argument. 
\end{Definition}

%This has been re-stated on several occasions~\cite{jiz:ari:04,gol:pas:yar:09}, to generate the \emph{R\'enyi entropies} of~\eqref{eq:RenyE}, and
R\'enyi also used the postulate approach to define the following quantity:
\begin{Definition}[\cite{ren:61,ren:70}]
The \emph{gain of information} or \emph{divergence (between distributions)} when $Y \sim P_Y$, $P_Y(y_i) = q_i$ is substituted by $X \sim P_X$, $P_X(x_i) = p_i$ being continuous wrt the latter---that is, with $\supp{Y} \subseteq \supp{X}$---as\footnote{ Such special cases will not be stated again, as motivated in Section~\ref{sec:rentropy:shifted}.}
%On a similar basis, R\'enyi defined the \emph{gain of information} 
%of \eqref{eq:RenyD} when substituting from distribution $Q$ by distribution $P$~\cite{ren:61,ren:70}.
%\todo{Restate R\'enyi's divergence.}
%{with associated divergence when $Q$ is substituted by $P$ on similar support $\mathcal X$ with $p_X$ and $q_X$}
%
\begin{align*}
%\label{eq:RenyD}
&D_{\alpha}( P_X \| P_Y) %= D_{\alpha}(p \| q) 
	= \frac{1}{\alpha-1}\log\sum_{i=1}^n p_i^\alpha q_i^{1- \alpha}\quad \alpha \neq 1 %\\
&\lim_{\alpha \rightarrow 1} D_{\alpha}( P_X \| P_Y) %= \lim_{\alpha \rightarrow 1}  D_{\alpha}(p \| q)   
&= D_{KL}(P_X \| P_Y)\,  %\notag
%	= \frac{1}{\alpha-1}\log\sum_{i=1}^np_i \left(\frac{p_i}{q_i}\right)^{\alpha -1}
%
\end{align*}
and the fact that Kullback-Leibler's divergence emerges as the limit when $\alpha \rightarrow 1$ follows from the same continuity argument as before.
\end{Definition}

As in the Shannon entropy case, the rest of the quantities arising in Information Theory can be defined in terms of the generalized entropy and its divergence~\cite{jiz:ari:04,erv:har:14}.

%% file: probRRVV.tex
%Let $(\mathcal A, \Sigma_{\mathcal A}, P)$ be a probability space, with
%$\mathcal A = \{ a_1, \ldots, a_n\}$ the set of outcomes of a random experiment, 
%$\Sigma_{\mathcal A}$ the sigma-algebra of this set and 
%probability measure $P: \mathcal A \rightarrow \nnR\,, P(a_i) = p_i, 1 \leq k \leq n$\,.
Shannon and R\'enyi set out to find how much information can be gained \emph{on average} by a single performance of an experiment $\Omega$ under different suppositions. 
For that purpose, let $(\Omega, \Sigma_{\Omega}, P)$ be a measure space, with
$\Omega= \{ \omega_1, \ldots, \omega_n\}$ the set of outcomes of a random experiment, 
$\Sigma_{\Omega}$ the sigma-algebra of this set and 
measure $P: \Omega \rightarrow \nnR\,, P(\omega_i) = p_i, 1 \leq k \leq n$\,. 
We define  the \emph{support of $P$}, as the set of outcomes with positive probability $\supp{P} = \{ \omega \in \Omega \mid P(\omega) > 0 \}$\,.

%\subsubsection{Random variables}
Let $(\mathcal X, \Sigma_{\mathcal X})$ be a measurable space with $\mathcal X$ a domain and $\Sigma_{\mathcal X}$ its sigma algebra and 
consider the random variable $X: \Omega \rightarrow \mathcal X$\,, that is, a measurable function so that 
for each set of $B \in \Sigma_{\mathcal X}$ we have $X^{-1}(B) \in \Sigma_\Omega$. 
Then $P$ induces a measure $P_X$ on $(\mathcal X, \Sigma_{\mathcal X})$ with $\forall x \in \Sigma_{\mathcal X},\, P_X(x) = P(X=x) = P(X^{-1}(x))$, where  $x$ is an event in $\Sigma_{\mathcal X}$, and $P_X(x) = \sum_{\omega_i \subseteq X^{-1}(x)} P(\omega_i)$\,
whereby $(\mathcal X, \Sigma_{\mathcal X}, P_X)$ becomes a measure space. 
We will use mostly $X \sim P_X$  to denote a random variable,  instead of its measurable space. 
The reason for this is that since information measures are defined on distributions, this is the more fundamental notion for us. 

 Sometimes co-occurring random variables are defined on the same sample space and sometimes on different ones. 
 Hence,  we will need another measure space sharing the same measurable space $(\Omega, \Sigma_{\Omega})$ but different measure, $(\Omega, \Sigma_{\Omega}, Q)$ with $Q(\omega_i) = q_i$\,. 

\begin{Remark}
Modernly, discrete distributions are sets or vectors of non-negative numbers adding up to $1$, but R\'enyi developed his theory for ``defective distributions'', that is, with $\sum_i P(\omega_i) \neq 1$ which are better described as ``positive measures''. 
In fact, we do not need to distinguish whether $P$ is a probability measure in the $(n-1)$-simplex $P \in \Delta^{n-1} \Leftrightarrow \sum_i P(\omega_i) = 1$ or in general a measure  $P \in \nnR^{n}$ and 
nothing precludes using the latter to define entropies---while it provides a bit of generalization this is the road we will take below (see~\cite{ren:61,ren:70} on using incomplete distributions with $\sum_i p_i < 1$). %Section~\ref{sec:renyi:entropy}, and
\end{Remark}

%% file: shiftedRenyiSemifields.tex
%\subsubsection{The restatement of the postulate approach}
%\subsubsection{The shifted R\'enyi entropy}
To leverage the theory of generalized means to our advantage, 
we start with a correction to R\'enyi's entropy definition: 
The investigation into the form of the transformation function for the R\'enyi entropy~\eqref{eq:exponential} is arbitrary in the parameter $\alpha$ that it chooses. 
In fact,  we may substitute in $r = \alpha - 1$ to obtain the pair of formulas:
\begin{align}
\label{eq:adapting}
\varphi'(h) &= b^{-rh}
&
\varphi'^{-1}(p) &= \frac{-1}{r}\log_b p  
\end{align}
\begin{Definition}
%In such case, for the basic information function $I(p_i) = -\log p_i$ 
%from the formula for information~\eqref{eq:KNentropy} we obtain 
The \emph{shifted R\'enyi entropy of order $r\neq 0$} for a discrete random variable $ X \sim P_X$, is the Kolmogorov-Nagumo  $\varphi'$-mean~\eqref{eq:KNentropy} of the information function $\hinf{p} = - \ln p$ over the probability values. 
\begin{align}
\label{def:srenyiE}
\srentropy{P_X} &=\frac{-1}{r}\log_b\left(\sum_i \frac{p_i}{\sum_k p_k} {p_i^{r}}\right)
&
\lim_{r\rightarrow 0} \srentropy{P_X}  &=  \sentropy{P_X}\,.
\end{align}
\noindent
Note that: 
\begin{itemize}
\item For $r\neq 0$ this is motivated by:
\begin{align*}
%\pentropy{X} &= 
\srentropy{P_X} 
	&= \frac{-1}{r}\log_b\left(\sum_i \frac{p_i}{\sum_k p_k} b^{r\log_b p_i}\right)
	= \frac{-1}{r}\log_b\left(\sum_i \frac{p_i}{\sum_k p_k} b^{\log_b p_i^r}\right)
	= \frac{-1}{r}\log_b\left(\sum_i \frac{p_i}{\sum_k p_k} {p_i^{r}}\right)\,.
%	= - \log\left(\sum_i p_i p_i^{r}\right)^\frac{1}{r}
%	= - \log \hmean{P_X}{P_X}
\end{align*}

\item For $r=0$ we can use the linear mean $\varphi(h) = ah + b$ with inverse $\varphi^{-1}(p)= \frac{1}{a}(p - b)$ as per the standard definition, leading to Shannon's entropy.  

\end{itemize}
\qed
\end{Definition}

\begin{Remark}
\label{rem:base}
The base of the logarithm is not important as long as it is maintained in $\varphi'(\cdot)$, $\hinf{\cdot}$ and their inverses, hence we leave it implicit. 
For some calculations---e.g. the derivative below---we explicitly provide a particular basis---e.g. $\log_e x = \ln x$. 
\end{Remark}

The shifted  divergence can be obtained in the same manner---the way that R\'enyi followed himself%--- using the Kolmogorov-Nagumo generalized $\varphi'$-mean. 
~\cite{ren:70}. 
\begin{Definition}
The shifted R\'enyi divergence between two distributions $P_X(x_i) = p_i$ and $Q_X(x_i) = q_i$ with compatible support is the following quantity. 
\begin{align}
\label{def:srenyiD}
\srdiv[r]{P_X}{Q_X} &= \frac{1}{r}\log \sum_i \frac{p_i}{\sum_k p_k} \left( \frac{p_i}{q_i}\right)^r
&
\lim_{r\rightarrow 0} \srdiv{P_X}{Q_X}  &=  \sdiv{P_X}{Q_X}\,.
\end{align}
\end{Definition}

Of course, the values of the R\'enyi entropy and divergence are not modified by this shifting. 
%The procedure above returns the same concept as the standard R\'enyi entropy:
%The first thing to prove is the equality of the definitions:
\begin{Lemma}
%Let $r = \alpha - 1$. Then the 
The R\'enyi entropy and the shifted R\'enyi entropy produce the same value, and similarly for their respective divergences.
\end{Lemma}
\begin{proof}
if we consider a new parameter $r = \alpha  - 1$ we have:
\begin{align*}
 \rentropy{P_X}
&= \frac{1}{1-\alpha}\log\left(\sum_{i=1}^n \frac{p_i^\alpha}{\sum_k p_k} \right) 
	= \frac{-1}{r}\log \left (\sum_{i=1}^n \frac{p_i^{r+1}}{\sum_k p_k} \right )
%\intertext{whence}
%\tilde H_p(P_X) &= 
%	= \frac{-1}{r}\log \left(\sum_{i=1}^n p_i p_i^{r}\right) 
	= - \frac{1}{r}  \log \left(\sum_{i=1}^n \frac{p_i}{\sum_k p_k}  p_i^{r}\right) = \pentropy{P_X}\,.
%\intertext{Introducing the weighted power mean of \eqref{eq:holder:wmean} we have}
%H_\alpha(P_X) &= - \log \hmean{P_X}{P_X} = \pentropy{P_X}
\end{align*}
%Introducing the weighted H\"older mean of \eqref{eq:holder:wmean} we have \eqref{eq:sRenyE}, 
and similarly for the divergence:
\begin{align*}
\rdiv{P_X}{Q_X}  
 &= \frac{1}{\alpha-1}\log\sum_{i=1}^n \frac{p_i^\alpha q_i^{1- \alpha} }{\sum_k p_k} = \frac{1}{r}\log\sum_{i=1}^n \frac{p_i^{r+1} q_i^{-r}} {\sum_k p_k} 
	= \frac{1}{r} \log \sum_{i=1}^n \frac{p_i}{\sum_k p_k}  \left(\frac{p_i }{q_i}\right)^{r}  %\\
	= \srdiv{P_X}{Q_X}
\end{align*}
The Shannon entropy and Kullback-Leibler divergences are clearly the limit cases. 
%\todo{The shifted R\'enyi divergence.}
\end{proof}

\subsubsection{The case for shifting the R\'enyi entropy}

So what could be the reason for the shifting? First and foremost, it is a re-alignment with the more basic concept of generalized mean. 
\begin{Proposition}
\label{lemma:renyispectrum}
The Shifted R\'enyi Entropy and Divergence are logarithmic transformations of the generalized power means:
\begin{align}
\label{eq:sRenyE}
\srentropy{P_X} &= \log \frac{1}{\hmean{P_X}{P_X}}\\
\label{eq:sRenyD}
\srdiv{P_X}{Q_X} & = \log \hmean{P_X}{\frac{P_X}{Q_X}}
\end{align}
\end{Proposition}
\begin{proof}
Simple identification of \eqref{def:srenyiE} and \eqref{def:srenyiD} in the definition of power mean definitions~\eqref{eq:holder:wmean}. %of Property~\ref{propo:prop:hmean}. %\ref{propo:prop:hmean:4}
\end{proof}
% 
%\todo{Table with the alignment of the R\'enyi and shifted R\'enyi entropies and the Holder means.}
\noindent 
Table~\ref{tab:entropies} lists the shifting of these entropies and their relation both to the means and to the original R\'enyi definition in the parameter $\alpha$.  
\begin{table*}
\centering
\small
\begin{tabular}{|l|c|l|l|r|r|}
\hline
Mean name & Mean $\hmean[r]{\vec w}{\vec x}$& Shifted entropy $\pentropy[r]{P_X}$ & Entropy name & $\alpha$ & $r$\\
\hline
Maximum & $\max_i x_i$ & $\tilde H_{\infty} = -\log \max_i p_i$ & min-entropy & $\infty$ & $\infty$ \\
\hline
Arithmetic & $ \sum_i w_i x_i $ & $\tilde H_{1} =-\log \sum_i p_i^2$ & R\'enyi's quadratic & $2$ & $1$ \\
%Arithmetic & & & & \\
\hline
Geometric & $\Pi_i x_i^{w_i}$ & $\tilde H_{0} = -\sum_i p_i \log p_i$ & Shannon's & $1$ & $0$\\
\hline
Harmonic & $ (\sum_i w_i \frac{1}{x_i})^{-1}$ & $\tilde H_{-1} = \log n$  & Hartley's & $0$ & $-1$\\
\hline 
Minimum & $\min_i x_i$ & $\tilde H_{-\infty} = -\log \min_i p_i$ & max-entropy & $-\infty$ & $-\infty$\\
\hline
\end{tabular}
\caption{Relation between the most usual weighted power means, R\'enyi entropies and shifted versions of them.}
\label{tab:entropies}
\end{table*}

\begin{Remark}
It is no longer necessary to make the distinction between the case $r \rightarrow 0$---Shannon's---and the rest, since the means are already defined with this caveat. 
This actually downplays the peculiar features of 
%Perhaps the most striking effect is the fact that 
Shannon's entropy, arising from the geometric mean when $\sum_i p_i = 1$:
\begin{align*}
\srentropy[0]{p_x} &= \log \frac{1}{\hmean[0]{P_X}{P_X}} 
	= - \log \left(\prod_i p_i^{p_i} \right)
	=  - \sum_i p_i \log {p_i} 	
\end{align*}
However, the prominence of the Shannon entropy will emerge once again in the context of rewriting entropies in terms of each other (\S~\ref{sec:entropy:rewriting}). 
\end{Remark}

Since the means are properly defined for all $r \in [-\infty, \infty]$, $\srentropy{P_X}$ is likewise properly defined for all $r \in [-\infty, \infty]$---and therefore the non-shifted version with $\alpha = r  + 1$. 
This is probably the single strongest argument in favour of the shifting and motivates the following definition. 
%Therefore, %since $\hmean{P_X}{P_X}$ is a peculiar kind of H\"older mean, 
\begin{Definition}[The R\'enyi information spectrum]
\label{def:renyi:spectrum}
For fixed $P_X$ we will refer to $\srentropy{P_X}$ as its \emph{R\'enyi information spectrum over parameter $r$}. 
\end{Definition}

Also, some relationships between magnitudes are clarified in the shifted enunciation with respect to the traditional one, 
%\todo[inline]{Develop the following. }
%Furthermore,  
%it also clarifies 
for instance, the relation between the R\'enyi entropy and divergence. 
\begin{Lemma}
\label{lemma:rewriting:D}
The shifted formulation makes the entropy the self-information with a change of sign:
\begin{align}
\srentropy{P_X} &= \srdiv[-r]{P_{XX}}{P_X P_X}\,.
\end{align}
\end{Lemma}
\begin{proof}
$\srdiv[-r]{P_{XX}}{P_X P_X} = \srdiv[-r]{P_{X}}{P_X P_X} = \frac{-1}{r}\log \sum_i p_i \left(\frac{p_i}{p_i p_i}\right)^{-r} = \frac{-1}{r}\log \sum_i p_i \left(\frac{1}{ p_i}\right)^{-r} = \srentropy{P_X}$\,.
%By rewriting $\pdiv[-r]{P_{XX}}{P_X P_X}$.
%\begin{align*}
%\srdiv[-r]{P_{XX}}{P_X P_X} &= \srdiv[-r]{P_{X}}{P_X P_X} = \frac{-1}{r}\log \sum_i p_i \left(\frac{p_i}{p_i p_i}\right)^{-r} = \frac{-1}{r}\log \sum_i p_i \left(\frac{1}{ p_i}\right)^{-r} = \srentropy{P_X} 
%%&= \frac{-1}{r}\log \sum_i p_i \left(\frac{1}{ p_i}\right)^{-r}= \frac{-1}{r}\log \sum_i p_i { p_i}^{r} = \srentropy{P_X} 
%\end{align*}
\end{proof}
\noindent
Recall that in the common formulation, $\rentropy{P_X} = \rdiv[2-\alpha]{P_X}{P_X P_X}$~\cite{erv:har:14}. 

% % % % % % % % % % % % % % % % % % % % % % % % % % % % % % % % % % % % % % %
%\subsubsection{Properties of the shifted R\'enyi entropy and divergence}
% % % % % % % % % % % % % % % % % % % % % % % % % % % % % % % % % % % % % %

Another simplification is the fact that the properties of the R\'enyi entropy and divergence stem from those of the means, inversion and logarithm, a great simplification. 
%
%with the notorious following properties:
\begin{Proposition}[Properties of the R\'enyi spectrum of $P_X$]
\label{prop:prop:sE}
Let $r,s \in \mathbb R\cup\{\pm\infty\}$, and 
$P_X, Q_X \in \Delta^{n-1}$ where $\Delta^{n-1}$ %$\mathbb S (\mathcal X)$ 
is the simplex over the support $\supp{X}$, with cardinal $|\supp{X}| = n$. Then, 
\begin{enumerate}
\item \label{prop:prop:sE:p11} (Monotonicity) 
	The R\'enyi entropy is a non-increasing function of the order $r$.
\begin{align}
	s \leq r \Rightarrow \srentropy[s]{P_X} \geq \srentropy[r]{P_X} 
\end{align}

\item \label{prop:prop:sE:p12} (Boundedness) 
The R\'enyi spectrum $\srentropy{P_X}$ is bounded by the limits
\begin{align}
	\srentropy[-\infty]{P_X}  \geq \srentropy[r]{P_X} \geq \srentropy[\infty]{P_X} 
\end{align}

\item \label{prop:prop:sE:p2} %(Constant uniform entropy)
	The entropy of the uniform \emph{pmf} $U_X$ is constant over $r$\,.
\begin{align}
	\forall r \in \mathbb R\cup\{\pm\infty\}\,,  \srentropy[r]{U_X} = \log n%|\mathcal X|
\end{align}

\item \label{prop:prop:sE:p3} %( Hartley )
	The Hartley entropy ($r=-1$) is constant over the distribution simplex. 
\begin{align}	
	%\forall P_X \in {\mathbb S}(\mathcal X)\,, 
	\srentropy[-1]{P_X} = \log n %|\mathcal X|
\end{align}
\item \label{prop:prop:sE:p5} (Divergence from uniformity) 
%Call $U_X$ the uniform distribution on $\mathcal X$. Then the 
The divergence of any distribution $P_X$ from the uniform $U_X$ can be written in terms of the entropies as: 
\begin{align}
	\srdiv[r]{P_X}{U_X} = \srentropy{U_X} - \srentropy{P_X}\,.
\end{align}

\item \label{prop:prop:sE:p7} (Derivative of the shifted entropy) 
The derivative in $r$ of R\'enyi's $r$-th order entropy is
\begin{align}
\frac{d}{dr}\srentropy{P_X} = \frac{-1}{r^2}\srdiv[0]{\srdist[r]{P_X}}{P_X}
	= \frac{-1}{r}\log\frac{\hmean[0]{\srdist[r]{P_X}}{P_X}}{\hmean{P_X}{P_X}}\,,
\end{align}
where $\srdist[r]{P_X}=\left\{\frac{p_ip_i^r}{\sum_k p_k p_k^r}\right\}_{i=1}^n$ for $r \in \mathbb R\cup\{\pm\infty\}$ are the \emph{shifted escort distributions}.

\item \label{prop:prop:sE:p6}  (Relationship with the moments of $P_X$) 
The shifted R\'enyi Entropy of order $r$ is the logarithm of the inverse $r$-th root of the $r$-th moment of $P_X$\,.
\begin{align}
\label{eq:sre:pot}
\srentropy{P_X} &= -\frac{1}{r} \log E_{P_X}\{P_X^r\} = \log \frac{1}{\sqrt[r]{ E_{P_X}\{P_X^r\} }}
\end{align}
\end{enumerate}
\end{Proposition}
\begin{proof} 
Property~\ref{prop:prop:sE:p11} issues from Property~\ref{propo:prop:hmean}.\ref{prop:hmean:2} and Hartley's information function being order-inverting or antitone%a dual-order isomorphism from Corollary~\ref{coro:hartleys}
\footnote{Properties used in the following are referred to the Proposition they are stated in.}. 
Since the free parameter $r$ is allowed to take values in $[-\infty, \infty]$, Property~\ref{prop:prop:sE:p12} follows directly from Property~\ref{prop:prop:sE:p11}. 
With respect to Property~\ref{prop:prop:sE:p2}, we have, from $U_X = 1/|\supp{X}| = 1/n$ and  Property~\ref{propo:KN}.\ref{prop:knmean:8}:
\[
\pentropy{\frac{1}{n}} = - \log \hmean{\frac{1}{n}}{\frac{1}{n}} = -\log \frac{1}{n} = \log n\,.
\]
For Property~\ref{prop:prop:sE:p3} we have:
\[
\pentropy[-1]{P_X} = - \log (\sum_i p_i \cdot p_i^{-1})^{-1} = - \log (n)^{-1} = \log n
\]
While for Property~\ref{prop:prop:sE:p5},
\begin{align*}
	\pdiv[r]{P_X}{U_X} 
		&= \frac{1}{r}\log\left[\sum_i p_i\left(\frac{p_i}{u_i}\right)^r\right]
		= \frac{1}{r}\log\left[\sum_i p_i\left(\frac{p_i}{1/n}\right)^r\right] 
		= \frac{1}{r}\log\left[n^r \left(\sum_i p_i p_i^r\right) \right]\\
		 & = \log n + \log \left(\sum_i p_i p_i^r\right)^{1/r} 
		= \pentropy{U_X} - \pentropy{P_X}\,.
\end{align*}
For the third term of Property~\ref{prop:prop:sE:p7}, we have from~\eqref{eq:sRenyE} with natural logarithm, with $P_X$ in the role both of $\vec w$ and $\vec x$\,
\begin{align*}
\frac{d}{dr}\pentropy{P_X} = \frac{d}{dr}\left( - \log_e \hmean[r]{P_X}{P_X} \right)= - \frac{\frac{d}{dr}\hmean[r]{P_X}{P_X}}{\hmean[r]{P_X}{P_X}}\,, 
\end{align*}
whence the property follows directly from~\eqref{eq:mean:derivative}.
%s
For the first identity, though, we have: %of Property~\ref{prop:prop:sE:p7}
\begin{align}
\frac{d\pentropy{P_X}}{dr} 
&= - \frac{d}{dr}  \left[\frac{1}{r}\log\sum_i p_i p_i^r\right] 
= - \left[ \frac{-1}{r^2}\log\sum_i p_i p_i^r + \frac{1}{r}\sum_i \frac{p_i p_i^r }{\sum_i p_i p_i^r} \log p_i\right]\,. \notag
\intertext{If we introduce the abbreviation}
\label{eq:def:escort}
\srdist[r]{P_X}&=\srdist[r]{P_X,P_X} = \{\srdist[r]{P_X}_i\}_{i=1}^n = \left\{\frac{p_ip_i^r}{\sum_k p_k p_k^r}\right\}_{i=1}^n
\intertext{noticing that $\log\sum_k p_k p_k^r  = \sum_i \srdist[r]{P_X}_i  \log (\sum_k p_k p_k^r)$, since $\srdist[r]{P_X}$ is a distribution, 
and factoring out $-1/r^2$:}
\frac{d\pentropy{P_X}}{dr} 
&= - \frac{1}{r^2}\left[ 
- \sum_i \srdist[r]{P_X}_i  \log (\sum_k p_k p_k^r) 
+ r\left( \sum_i \srdist[r]{P_X}_i  \log p_i  \right)
\pm \sum_i \srdist[r]{P_X}_i  \log p_i 
\right] \notag 
\\
&= - \frac{1}{r^2}\left[ 
- \sum_i \srdist[r]{P_X}_i  \log (\sum_k p_k p_k^r) 
+( r +1)\sum_i \srdist[r]{P_X}_i  \log p_i
- \sum_i \srdist[r]{P_X}_i  \log p_i 
\right]  \notag 
\\
&= - \frac{1}{r^2}\left[ 
 \sum_i \srdist[r]{P_X}_i  \log \frac{p_i p_i^r}{\sum_k p_k p_k^r} 
- \sum_i \srdist[r]{P_X}_i  \log p_i 
\right]  \notag 
%\\
= - \frac{1}{r^2} \sum_i \srdist[r]{P_X}_i  \log \frac{\srdist[r]{P_X}_i}{p_i} 
%= - \frac{1}{r^2} \srdiv[0]{\srdist[r]{P_X}}{P_X}
\end{align}
%\begin{align*}
%\frac{d\pentropy{P_X}}{dr} 
%&= \frac{d\pentropy{P_X}}{d\rentropy{P_X}} \cdot \frac{d\rentropy{P_X}}{d\alpha} \cdot \frac{dr}{d\alpha}
%\intertext{but since $a = r - 1$ and $\rentropy{P_X} = \pentropy{P_X}$}
%&= \frac{d\rentropy{P_X}}{d\alpha}
%\intertext{
and recalling the definition of the shifted divergence we have the result.
%} 
%&= - \frac{1}{(\alpha -1)^2}\rdiv[1]{\rdist{P_X}}{P_X} 
%= - \frac{1}{r^2}\pdiv[0]{\rdist[r]{P_X}}{P_X}\,.
%\end{align*}

For Property~\ref{prop:prop:sE:p6}, %Remember that a Random Variable is actually a measurable function of some random space. Consider a probability space $(\Omega, \Sigma_\Omega, \mathcal P)$ where $\Omega$ is a set of random events, $\Sigma_\Omega$ is a sigma algebra of such set and $\mathcal P$ is a probability measure defined over the events of $\Sigma_\Omega$\,. 
in particular, the \emph{probability of any event} is a function of the random variable $P_X(x_i)=p_i$ whose \emph{$r$-th moment of $P_X$} is
\begin{align}
\label{eq:pot:mean}
E_X\{P_X^r\} = \sum_i p_i p_i^r = \left(M_r(P_X, P_X)\right)^r
\end{align}
The result follows by applying the definition of the shifted entropy in terms of the means. 
%are the \emph{moments of the probability measure} itself. 
%
\end{proof}

\begin{Remark}
In the preceding proof we have introduced the notion of \emph{shifted escort probabilities} $\srdist[r]{P_X}$ acting in the shifted R\'enyi entropies as the analogues of the \emph{escort probabilities} in the standard definition (see \cite{bec:sch:95} and section~\ref{sec:means}). 
This notion of shifted escort probabilities  is the one requested by Property~\ref{propo:prop:hmean}.\ref{prop:hmean:6} 
%of Proposition~\ref{propo:prop:hmean} 
 by instantiation of variables $\srdist[]{P_X} = \srdist[]{P_X, P_X}$\,. 
 But notice also that $(\srdist{P_X})_i = \frac{p_i p_i^r}{\sum_k p_k p_k^r} = \frac{p_i^{\alpha}}{\sum_k p_k^{\alpha}}=(\rdist{P_X})_i$ %= (\rdist[r+1]{P_X})_i$, 
 is just the shifting of the traditional escort probabilities~\cite{bec:sch:95}. 
 
 Note that for $P_X \in \nnR^{n}$: 
 \begin{itemize}
\item  $\srdist[0]{P_X}$ is the normalization of $P_X$. In fact,  $P_X \in \Delta^{n-1}$ if and only if we have $\srdist[0]{P_X} = P_X$\,. 
 \item 
 %for $P_X \in \nnR^{n}$, then 
 $\srdist[-1]{P_X}(x_i)= |\supp{P_X}|^{-1}$ if $x_i \in \supp{P_X}$ and $0$ otherwise. 
\item Furthermore,  if $P_X$ has $P$ maxima ($p$ minima), then $\srdist[\infty]{P_X}$  ($\srdist[-\infty]{P_X}$) is a distribution null everywhere but at the indices where the maxima (minima) of $P_X$ are situated:
 \begin{align*}
 \srdist[\infty]{P_X}(x_i) &= 
 \begin{cases}
 \frac{1}{P} & x_i \in \arg \max P_X\\
 0 & \text{otherwise}
 \end{cases}
 &
  \srdist[-\infty]{P_X}(x_i) &= 
  \begin{cases}
  \frac{1}{M} & x_i \in \arg \min P_X\\
  0 & \text{otherwise}
  \end{cases}
 \end{align*}
%\noindent Finally, $P_X \in \Delta^{n-1}$ if and only if we have $\srdist[0]{P_X} = P_X$\,. 
 \end{itemize}
\end{Remark}

%, not just the probability mass functions is made also clear. 

%\subsubsection{R\'enyi entropies on non-probabilistic measures}
\label{sec:Renyi:measures}
Another important point made clear by this relation to the means is the fact that \emph{all positive measures have a R\'enyi spectrum}: 
although so far we conceived the origin of information to be a probability function, nothing precludes applying the same procedure to non-negative, non-normalized quantities with $\sum_x f_X(x) \not = 1$, e.g. masses, sums, amounts of energy, etc. 

It is well-understood that in this situation Renyi's entropy has to be slightly modified to accept this procedure. The reason for this is Property~\ref{propo:prop:hmean}.\ref{prop:hmean:1}  of the means: generalized means are $1$-homogeneous in the numbers being averaged, but $0$-homogeneous in the weights. In the R\'enyi spectrum both these roles are fulfilled by the pmf.
% and this entails that 
Again the escort distributions allow us to analyze the measure: 
\begin{Lemma}
\label{propo:sRenyi:measures}
Consider a random variable $X \sim M_X$ with non-normalized measure $M_X(x_i) = m_i$ such that $\sum_i m_i = M \neq 1$\,. Then the normalized probability measure $\srdist[0]{M_X} = \{ m_i / \sum_i m_i \} _{i=1}^n $ provides a R\'enyi spectrum that is displaced relative to that of the measure as: 
\begin{align}
\label{eq:mass:edispl}
\srentropy[r]{M_X}  = \srentropy[r]{\srdist[0]{M_X}} - \log M\,.
\end{align}
\end{Lemma}
\begin{proof}
\begin{align*}
\srentropy[r]{\srdist[0]{M_X}} &= - \log \hmean[r]{\srdist[0]{M_X}}{\srdist[0]{M_X}} = - \frac{1}{r} \log \sum_i \frac{m_i}{M} \left ( \frac{m_i}{M} \right )^r 
= \log M - \frac{1}{r} \log \sum_i \frac{m_i}{M} m_i^r 
\\
&= \log M - \log \hmean[r]{M_X}{M_X}   = \log M + \srentropy[r]{M_X} 
\end{align*}
%so the entropy spectrum of the mass function is displaced by an amount $-\log M$ with respect to that of its probability distribution:
\end{proof}

\begin{Remark}
When $M \geq  1, -\log M \leq 0$ with equality for $M=1$ and that if $M < 1$ then $-\log M > 0$. 
This last was the original setting R\'enyi envisioned and catered for in the definitions, but nothing precludes the extension provided by Lemma~\ref{propo:sRenyi:measures}. 
In this paper, although $P_X$ can be interpreted as a \emph{pmf} in the formulas, it can also be interpreted as a mass function as in the Lemma above. However, the escort probabilities are always \emph{pmf}s. 
\end{Remark}

\begin{Example}
This example uses the UCB admission data from~\cite{bic:ham:oco:75}. 
We analyze the distribution of admissions with count vector $M_X = [933\; 585\; 918\; 792\; 584\; 714]^\top$
and probabilities $\srdist[0]{M_X}  \approx [0.21\; 0.13\; 0.20\; 0.17\; 0.13\; 0.16]^\top$. 
%We analyze the distribution of admissions by department with count vector $M_X = [933\; 585\; 918\; 792\; 584\; 714]^\top$
%and probabilities $P_X \approx [0.21\; 0.13\; 0.20\; 0.17\; 0.13\; 0.16]^\top$. 
%
The names of the departments are not important, due to the symmetry property. 
Figure~\ref{fig:eq:prob}.\subref{fig:eq:prob:RS} shows the R\'enyi Spectrum extrapolated from a sample of some orders which include $r \in \{-\infty, -1, 0, 1, \infty\}$. 
\qed
\begin{figure}
\begin{subfigure}[b]{0.5\textwidth}
%\subfloat[R\'enyi spectrum of $P_X$]{
\includegraphics[scale=0.1]{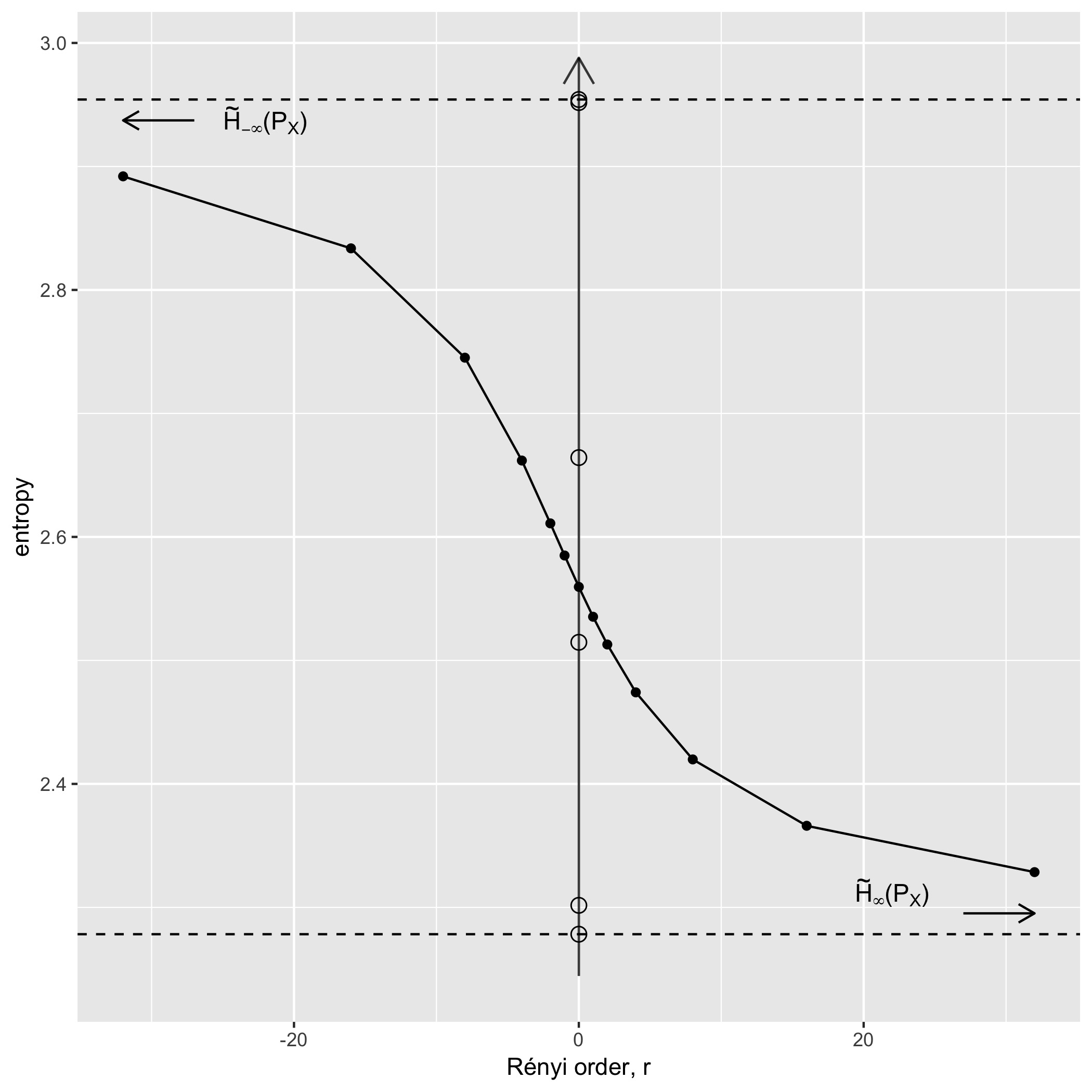}
\caption{R\'enyi spectrum of $\srdist[0]{M_X}$}
\label{fig:eq:prob:RS}
\end{subfigure}
\begin{subfigure}[b]{0.5\textwidth}
%\subfloat[Equivalent probability function aka H\"older path of $P_X$]{
%%\hspace{7cm}
\includegraphics[scale=0.1]{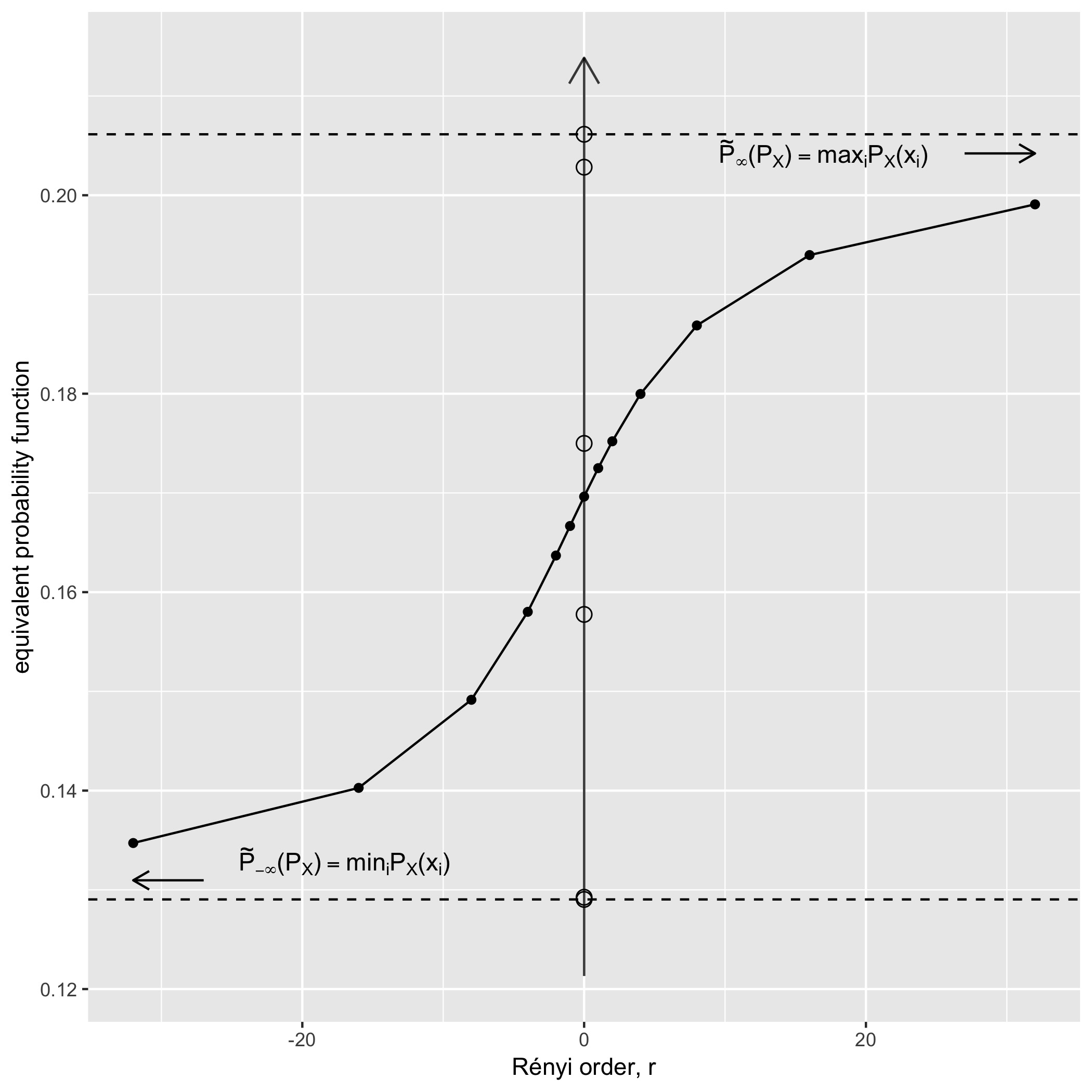}
\caption{Equivalent probability function of $\srdist[0]{M_X}$}
\label{fig:eq:prob:EPF}
\end{subfigure}
\caption{\textbf{R\'enyi spectrum (left) and equivalent probability function (right, also H\"older path, see \S~\protect\ref{sec:ca:rentropy})  of $\srdist[0]{M_X}$, the probability distribution of a simple mass measure $M_X$ with $n=6$.}
The values of the self-information (left) and probability (right) of the original distribution are shown at $r=0$ also (hollow circles): only 5 values seem to exist because the maximal information (minimal probability) is almost superposed on a second.}
\label{fig:eq:prob}
\end{figure}
%\todo[inline]{Include plot of R\'enyi spectrum and equivalente probabilities using the sigmoids.}
\end{Example}
%%\todo{The peculiarities of the Hartley entropy}
%When $\alpha = 0$ we have $r=-1$, that is, we consider the Hartley entropy\cite{keylist}. 
%Note that in the Hartley entropy, any distribution would have an entropy similar to that of the uniform, by Properties~\ref{prop:prop:sE}.\ref{prop:prop:sE:p2}  and~\ref{prop:prop:sE}.\ref{prop:prop:sE:p3}.
%But also, that two distributions are always ``independent''.

%Notice that:
%\begin{itemize}
%%\item The $\Delta H_{XY}$ coordinate of the Entropy Triangle is actually a divergence,
%%\[
%%\Delta H_{XY} = \pentropy[0]{U_X} - \pentropy[0]{P_X} = \pdiv[0]{P_X}{U_X}\,.
%%\]
%
%\item We could have proven Property~\ref{prop:prop:sE:p5} directly from Property~\ref{prop:hmean:7}, what strengthens the alignment of the shifted entropies with the power means. 
%
%\item Alternatively, the antitone character of the shifted entropies in $r$, Property~\ref{prop:prop:sE:p1} can be proven from Property~\ref{prop:prop:sE:p5}.
%\end{itemize}

%\todo[inline]{Review the definition of the Cross-entropy and its relation to the dvivergence.}
%\subsubsection{The Shifted R\'enyi Cross-Entropy}
%\input{shiftedCrossentropy.tex}

\subsubsection{Shifting other concepts related to the entropies}
Other entropy-related concepts may also be shifted. In particular, the cross-entropy has an almost direct traslation. 
\begin{Definition}
The shifted R\'enyi cross-entropy of order $r \in [-\infty, \infty]$ between two distributions $P_X(x_i) = p_i$ and $Q_X(x_i) = q_i$ with compatible support is
\begin{align}
\label{eq:sxent}
\srcrossent{P_X}{Q_X}  &= \log \frac{1}{\hmean{P_X}{Q_X}} %= - \frac{1}{r}\log  \sum_i p_i \left( \frac{p_i}{q_i}\right)^r
%\srcrossent{P_X}{Q_X} & = - \log \hmean{P_X}{Q_X} = \frac{-1}{r}\log\sum_i \frac{p_i}{\sum_k p_k} q_i^r & r &\neq 0
%\\
%\srcrossent[0]{P_X}{Q_X} &= - \log \hmean[0]{P_X}{Q_X}= - \sum_i \frac{p_i}{\sum_k p_k} \log q_i  & r &= 0 \notag
\end{align}

\end{Definition}
Note that the case-based definition is redundant: %, e.g. $\srcrossent{P_X}{Q_X}  = - \log \hmean{P_X}{Q_X}, \forall r \in \mathbb R$. 
the Shannon cross-entropy appears as $\srcrossent[0]{P_X}{Q_X} = - \log \hmean[0]{P_X}{Q_X}= - \sum_i \frac{p_i}{\sum_k p_k} \log q_i$, while for $r \neq 0$ we have $\srcrossent{P_X}{Q_X} = - \frac{1}{r}\log  \sum_i p_i \left( \frac{p_i}{q_i}\right)^r$ by virtue of the definition of the means again. 

Perhaps the most fundamental magnitude is the cross-entropy since it is easy to see that:
\begin{Lemma}
\label{lemma:crosse:basis}
In the shifted formulation both the entropy and the divergence are functions of the cross-entropy:
\begin{align}
\label{eq:crosse:basis}
\srentropy{P_X} &= \srcrossent{P_X}{P_X}
&
\srdiv{P_{X}}{Q_X} &= \srcrossent[-r]{P_X}{Q_X/P_X}
\end{align}
\end{Lemma}
\begin{proof}
The first equality is by comparison of definitions, while the second comes from:
\begin{align*}
%\srdiv[-r]{P_{X}}{Q_X} &= \frac{1}{-r}\log\sum_i p_i \left( \frac{p_i}{q_i}\right)^{-r}
%			= - \frac{1}{r}\log\sum_i p_i \left( \frac{q_i}{p_i}\right)^{r} %\\
% 			= \srcrossent{P_X}{Q_X/P_X}\,.
\srdiv{P_{X}}{Q_X} &= \frac{1}{r}\log\sum_i p_i \left( \frac{p_i}{q_i}\right)^{r} 
	= - \frac{1}{-r}\log\sum_i p_i \left( \frac{q_i}{p_i}\right)^{-r} 
	= \srcrossent[-r]{P_X}{Q_X/P_X}
\end{align*}
%Similar to Lemma~\ref{lemma:rewriting:D}, by an easy rewriting.
\end{proof}
%Note that the expression 
Note that if we accept the standard criterion in Shannon's entropy $0 \times \log \frac{1}{0} = 0 \times \infty = 0$ then  the previous expression for the cross-entropy is defined even if $p_i = 0$.

%% file: rewritingRE.tex
%\begin{Remark}
Not every expression valid in the case of Shannon's entropies can be translated into R\'enyi entropies: 
recall from the properties of the Kullback-Leibler divergence its expression in terms of the Shannon entropy and cross-entropy. We have: 
\begin{align}
\label{eq:shan:div:cross}
\srdiv[0]{P_X}{Q_X} &= - \srentropy[0]{P_X} + \srcrossent[0]{P_X}{Q_X}, 
\end{align}
but, in general, $\srdiv{P_X}{Q_X} \neq - \srentropy{P_X} + \srcrossent{P_X}{Q_X}$. %(however, see section~\ref{sec:entropy:rewriting}). 
%\end{Remark}

However, the shifting sometimes helps in obtaining ``derived expressions''. 
In particular, the (shifted) escort probabilities are ubiquitous in expressions dealing with R\'enyi entropies and divergences, and  allow us to discover the deep relationships between their values for different $r$'s. 
\begin{Lemma}
\label{lemma:ren:shan}
Let $r,s \in \mathbb R\cup\{\pm\infty\}$, 
$P_X \in \Delta^{n-1}$ where $\Delta^{n-1}$ is the simplex over the support $\supp{X}$. Then, 
\begin{align}
\label{eq:ren:shan:div}
\srentropy{P_X} &= \frac{1}{r}\srdiv[0]{\srdist{P_X}}{P_X}  + \srcrossent[0]{\srdist{P_X}}{P_X}
\\
\label{eq:ren:shan:ent}
\srentropy{P_X} &= \frac{-1}{r}\srentropy[0]{\srdist{P_X}}  + \frac{r+1}{r}\srcrossent[0]{\srdist{P_X}}{P_X}
\end{align}
\end{Lemma}
\begin{proof}
First, from the definitions of shifted R\'enyi entropy and cross-entropy and Property~\ref{prop:prop:sE}.\ref{prop:prop:sE:p7} we have:
\begin{align*}
\frac{-1}{r^2} \srdiv[0]{\srdist{P_X}}{P_X} = \frac{1}{r}\left[  \srentropy{P_X}  - \srcrossent[0]{\srdist{P_X}}{P_X}\right]
\end{align*}
Solving for $\srentropy{P_X}$ obtains the first result. 
By applying~\eqref{eq:shan:div:cross} to $\srdist{P_X}$ and $P_X$ we have:
\begin{align}
\label{eq:snd:rel}
\srdiv[0]{\srdist{P_X}}{P_X} &= - \srentropy[0]{\srdist{P_X}} + \srcrossent[0]{\srdist{P_X}}{P_X}\,.
\end{align}
and putting this into \eqref{eq:ren:shan:div} obtains the second result.

Another way is to prove it is from the definition of 
%First, from the definition of shifted R\'enyi entropy and escort probability, we have:
\begin{align*}
%\label{eq:first:rel}
\srentropy[0]{\srdist{P_X}} &= - \sum_i \frac{p_i p_i^r}{\sum_k p_k p_k^r} \log \frac{p_i p_i^r}{\sum_k p_k p_k^r}
= \sum_i \srdist{P_X} \log \left( \sum_k p_k p_k^r  \right) - \sum_i \srdist{P_X} \log p_i^{r+1} =
\notag \\
& = \log \left( \sum_k p_k p_k^r  \right)  - (r+1) \sum_i \srdist{P_X} \log p_i %=
%\notag \\
%& 
=  -r \srentropy{P_X} + (r+1) \srcrossent[0]{\srdist{P_X}}{P_X}
\end{align*}
and reorganize to obtain \eqref{eq:ren:shan:ent}. 
Again inserting the definition of the Shannon divergence in terms of the cross-entropy \eqref{eq:snd:rel},  into \eqref{eq:ren:shan:ent} and reorganizing  we get \eqref{eq:ren:shan:div}.
\end{proof}

On other occasions, using the shifted version does not help in simplifying expressions. For instance \emph{skew symmetry} looks in the standard case as $\rdiv{P_X}{Q_X} = \frac{\alpha}{1-\alpha}\rdiv[1-\alpha]{Q_X}{P_X}$, for any $0 < \alpha < 1$~\cite[Prop.~2]{erv:har:14}. In the shifted case we have the slightly more general expression for $r \neq 0$:
\begin{Lemma}
When $Q\sim Q_X$ is substituted by $P\sim P_X$ on a compatible support, then:
\begin{align}
\srdiv{P_X}{Q_X} = - \frac{r+1}{r}\cdot\srdiv[-(r+1)]{Q_X}{P_X}
\end{align}
\end{Lemma}
\begin{proof}
By easy manipulation of the definition of the divergence. 
\end{proof}

%% file: circaRentropy.tex
On the one hand, the existence of Hartley's information function \eqref{eq:Hartley_s} ties up all information values to probabilities and \emph{vice-versa}. On the other, R\'enyi's averaging function and its inverse \eqref{eq:adapting} also transform probabilities into informations and \emph{vice-versa}. 
In this section we explore the relationship between certain quantities generated by these functions, probabilities and entropies. 

\subsubsection{The equivalent probability function}
\label{sec:rentropy:prob}
\input{equivalentProb.tex}

\subsubsection{The information potential}
\label{sec:rentropy:part}
\input{partitionPotential.tex}

\subsubsection{Summary}
\label{sec:rentropy:sum}
Table~\ref{tab:circaRE} offers a summary of the quantities mentioned above and their relationships, while  the domain diagram in Figure~\ref{fig:circaREntropy} %.\subref{fig:circaREntropy:magnitudes} 
summarizes the actions of these functions to obtain the shifted R\'enyi entropy. 
A similar diagram is, of course, available for the standard entropy, using $\varphi$ with the $\alpha$ parameter. 
\begin{figure}
%\begin{minipage}{0.5\linewidth}
\begin{subfigure}[b]{0.5\textwidth}
	\centering
%\subfloat[Between entropy-related quantities]{
	\begin{tikzpicture}[scale=1.0]
	\node (P)  at (3,2) {$\hmean{P_X}{P_X}$};
	\node (H) at (0,0) {$\srentropy{P_X }$};
	\node (V)  at (3,-2) {$\srpot{P_X}$};
	\path[->,font=\scriptsize,>=angle 90]
	(P) edge  [bend left=10] node[below right]{$\mathfrak I^\ast$} (H)
	(H) edge  [bend left=10] node[above right]{$\varphi'$} (V)
	(V) edge  [bend left=10] node[left]{$\cdot^{r}$} (P)
	(P) edge [bend left=10] node[right]{$\cdot^{1/r}$} (V)
	%(V) edge [loop right] node[right] {$\langle \cdot \rangle_{P_{X}}$} (P_r)
	(V) edge [bend left=10] node[below left]{$\varphi'^{-1}$} (H)
	(H) edge [bend left=10] node[above left ]{${\left(\mathfrak I^\ast\right)}^{-1}$} (P)
	;
	%\node (A) at (0,2) {$\phantom{\quad}\mathcal X\phantom{\quad}$}; % \cong \rnorm[\gamma]{{\mathcal K}^g}$};
	%\node (B) at (3,2) {$\phantom{\quad}\mathcal Y\phantom{\quad}$};
	%\node (C) at (0,0) {$ \CL^{\gamma}_G(G,M,R)_{\mathcal K}$}; %=\extentOI{R}{\left(\rnorm[\mu]{\mathcal Y} \right)}$};
	%\node (D) at (3,0) {$ \CL^{\mu}_M(G,M,R)_{\mathcal K}$}; %=\extentOI{R}{\left(\rnorm[\mu]{\mathcal Y} \right)}$};
	%\path[->,font=\scriptsize,>=angle 90]
	%(A) edge [bend left=10] node[above]{$\intentOI{R}{\cdot}$} (B)
	%	edge [bend right=10] node[left]{$\pi_{\transp R}$} (C)
	%	edge node [near start, above, sloped] {$\intentOI{R}{\cdot}$} (D)
	%(B) edge [bend left=10] node[below]{$\extentOI{R}{\cdot}$} (A)
	%	edge [bend right=10] node[left]{$\pi_{R}$} (D)
	%	edge node [near start, below, sloped] {$\extentOI{R}{\cdot}$} (C)
	%(C) edge [bend left=10] node[above]{$\intentOI{R}{\cdot}$} (D)
	%	edge [bend right=10] node[right]{$\hookrightarrow_{\mathcal X}$} (A)
	%(D) edge [bend left=10] node[below]{$\extentOI{R}{\cdot}$} (C)
	%	edge [bend right=10] node[right]{$\hookrightarrow_{\mathcal Y}$} (B)
	%; %Don't forget this semicolon at the end of the path or tikz will refuse to compile!
	\end{tikzpicture}
	\caption{Between entropy-related quantities}
	\label{fig:circaREntropy:magnitudes}
%}
%\end{minipage}%
\end{subfigure}
%\hfill %Too much space apart
%\hspace{4cm}
\begin{subfigure}[b]{0.5\textwidth}
%\begin{minipage}{0.5\linewidth}
	\centering
%\subfloat[Between entropy-related domains (see \S~\ref{sec:entropy:smf})]{
	\begin{tikzpicture}[scale=1.0]
	\node (P)  at (3,2) {$\nnR$};
	\node (H) at (0,0) {$\mathbb H$};
	\node (P_r)  at (3,-2) {$\nnR^r$};
	\path[->,font=\scriptsize,>=angle 90]
	(P) edge  [bend left=10] node[below right]{$\mathfrak I^\ast$} (H)
	(H) edge  [bend left=10] node[above right]{$\varphi'$} (P_r)
	(P_r) edge  [bend left=10] node[left]{$\cdot^{r}$} (P)
	(P) edge [bend left=10] node[right]{$\cdot^{1/r}$} (P_r)
	%(V) edge [loop right] node[right] {$\langle \cdot \rangle_{P_{X}}$} (P_r)
	(P_r) edge [bend left=10] node[below left]{$\varphi'^{-1}$} (H)
	(H) edge [bend left=10] node[above left ]{${\left(\mathfrak I^\ast\right)}^{-1}$} (P)
	;
	%\node (A) at (0,2) {$\phantom{\quad}\mathcal X\phantom{\quad}$}; % \cong \rnorm[\gamma]{{\mathcal K}^g}$};
	%\node (B) at (3,2) {$\phantom{\quad}\mathcal Y\phantom{\quad}$};
	%\node (C) at (0,0) {$ \CL^{\gamma}_G(G,M,R)_{\mathcal K}$}; %=\extentOI{R}{\left(\rnorm[\mu]{\mathcal Y} \right)}$};
	%\node (D) at (3,0) {$ \CL^{\mu}_M(G,M,R)_{\mathcal K}$}; %=\extentOI{R}{\left(\rnorm[\mu]{\mathcal Y} \right)}$};
	%\path[->,font=\scriptsize,>=angle 90]
	%(A) edge [bend left=10] node[above]{$\intentOI{R}{\cdot}$} (B)
	%	edge [bend right=10] node[left]{$\pi_{\transp R}$} (C)
	%	edge node [near start, above, sloped] {$\intentOI{R}{\cdot}$} (D)
	%(B) edge [bend left=10] node[below]{$\extentOI{R}{\cdot}$} (A)
	%	edge [bend right=10] node[left]{$\pi_{R}$} (D)
	%	edge node [near start, below, sloped] {$\extentOI{R}{\cdot}$} (C)
	%(C) edge [bend left=10] node[above]{$\intentOI{R}{\cdot}$} (D)
	%	edge [bend right=10] node[right]{$\hookrightarrow_{\mathcal X}$} (A)
	%(D) edge [bend left=10] node[below]{$\extentOI{R}{\cdot}$} (C)
	%	edge [bend right=10] node[right]{$\hookrightarrow_{\mathcal Y}$} (B)
	%; %Don't forget this semicolon at the end of the path or tikz will refuse to compile!
	\end{tikzpicture}
	\caption{Between entropy-related domains}
	\label{fig:circaREntropy:domains}
\end{subfigure}

\caption[]{Schematics of relationship between some magnitudes in the text and their domains of definition (see \S~\ref{sec:semifields})}
\label{fig:circaREntropy}
\end{figure}
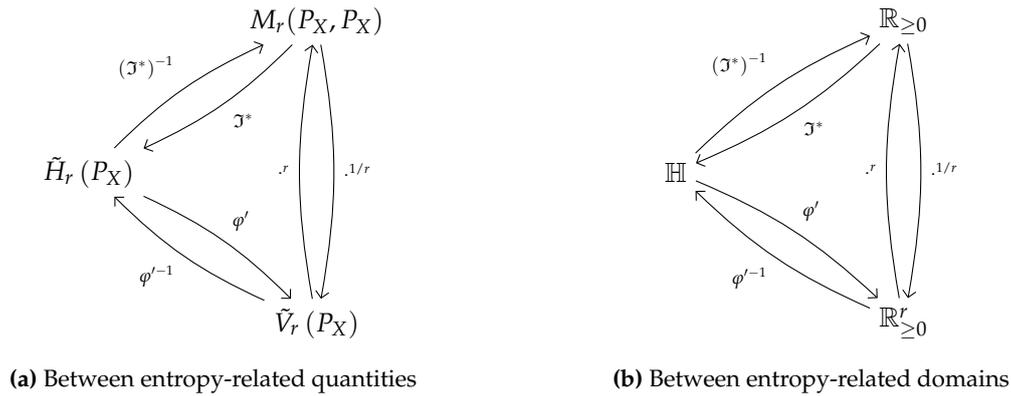
\begin{table*}
\begin{tabularx}{\textwidth}{|l||X|X|X||c|}
\hline 
quantity in terms of\ldots &  
R\'enyi entropy  & 
Gen. H\"older mean & 
Information potential &
distribution
%R\'enyi entropy \newline  $\srentropy{P_X}$ & 
%G. H\"older mean\newline  $\hmean{P_X}{P_X}$ & 
%partition function, expectation, information potential \newline
%$Z_r(P_X) = E_{P_X}\{P_X^r\} = V_r(P_X) $
\\
%& $\srentropy{P_X}$ & $\hmean{P_X}{P_X}$ & $\srpot{P_X}$ & $P_X = \{p_i\}_{i=1}^n$
%\\
\hline
\hline
R\'enyi entropy & $\srentropy{P_X }$ & $-\log \hmean{P_X}{P_X}$ & $\frac{-1}{r}\log \srpot{P_X}$ & $\frac{-1}{r}\log\left (\sum_i \frac{p_i}{\sum_k p_k} p_i^r\right)$
\\
\hline 
Gen. 
H\"older mean & $\exp(-\srentropy{P_X})$  & $\hmean{P_X}{P_X}$  & $\left (\srpot{P_X}\right)^\frac{1}{r}$ & $\left (\sum_i \frac{p_i}{\sum_k p_k} p_i^r\right)^\frac{1}{r}$
\\
\hline
Information potential &  $\exp(-r\srentropy{P_X})$   & $\hmean{P_X}{P_X}^r$  & $\srpot{P_X} = E_{P_X}\{P_X^r\}$
& $\sum_i \frac{p_i}{\sum_k p_k} p_i^r$
\\
\hline
\end{tabularx}
\caption{Quantities around the shifted R\'enyi entropy of a discrete distribution $P_X$.
\label{tab:circaRE}
}
\end{table*}

Note that these quantities have independent motivation: this is historically quite evident in the case of the means~\cite{har:lit:pol:52}, and the R\'enyi information~\cite{ren:61} and little bit less so in the case of the information potential which arose in the context of ITL~\cite{pri:10}, hence motivated by a desire to make R\'enyi's entropies more useful.
Both quantities are generated from/generate entropy by means by independently motivated functions, Hartley's transformation~\eqref{eq:Hartley_s}  and R\'enyi's transformation~\eqref{eq:adapting}, respectively.

Following the original  axiomatic approach it would seem we first transform the probabilities into entropies using Hartley's function and then we use the $\varphi'$ function to work out an average of these using the Kolmogorov-Nagumo formula. But due to the formulas for the information potential and the equivalent probability function we know that this is rather a composition of transformations, than a forward backward moving between entropies and probabilities. 
It is clear that the Hartley function and R\'enyi's choice of averaging function are special for entropies, from the postulate approach to their definition. %The next section provides another explanation of their importance. 

%% file: equivalentProb.tex
Recall that, due to Hartley's function, from every average measure of information, an equivalent \emph{average} probability emerges.
To see this in a more general light, first 
%Define 
define the extension to Hartley's information function to non-negative numbers $\hinf{\cdot}: [0,\infty] \rightarrow [-\infty, \infty]$ as $\hinf{p}= - \ln p$. This is  one-to-one from $[0,\infty]$ and total onto $[-\infty, \infty]$, with inverse $\hinfinv{h} = e^{-h}$ for $h \in [-\infty, \infty]$.  
\begin{Definition}
Let $X \sim P_X$ with R\'enyi spectrum $\srentropy[r]{P_X}$. 
Then the \emph{equivalent probability function of $\srprob[r]{P_X}$} is the Hartley inverse of $\srentropy[r]{P_X}$ over all values of $r \in [-\infty, \infty]$ 
\begin{align}
\label{eq:eq:prob}
\srprob[r]{P_X} = \hinfinv{\srentropy[r]{P_X}}
\end{align}
\end{Definition}

\begin{Remark}
\label{rem:epf}
The equivalent probability function for a fixed probability distribution $P_X$ is a function of parameter $r$---like the R\'enyi entropy---whose values are probabilities---in the sense that it produces values in $[0,1]$---but it is \emph{not} a probability distribution.  

Analogously, due to the extended definition of the Hartley information, this mechanism, when operating on a mass measure $M_X$, generates and \emph{equivalent mass function} $\srprob[r]{M_X}$, which is \emph{not} a mass measure. 
\end{Remark}

\begin{Lemma}
\label{lemma:epf:holder}
Let $X \sim P_X$. The \emph{equivalent probability function}  $\srprob[r]{P_X}$ is the H\"older path of the probability function  $P_X$  (as a set of numbers) using the same probability function as weights.
\begin{align}
\srprob[r]{P_X} = \hmean{P_X}{P_X}
\end{align}
\end{Lemma}
\begin{proof}
From the definition, using $b$ as the basis chosen for the logarithm in the information function.
\begin{align*}
\srprob[r]{P_X} = \hinfinv{\srentropy[r]{P_X}} = b^{-\srentropy[r]{P_X}} = b^{\log_b \hmean{P_X}{P_X}} = \hmean{P_X}{P_X}
\end{align*}
\end{proof}
%Note that this exceeds the range of the probability distributions and includes all sets of non-negative numbers. 

Note that by Remark~\ref{rem:epf} these means apply, in general,  to sets of non-negative numbers and not only to the probabilities in a distribution, given their homogeneity properties.
In the light of Lemma~\ref{lemma:epf:holder}, the following properties of the equivalent probability function are a corollary of those of the weighted generalized power means of Proposition~\ref{propo:prop:hmean} in  Section~\eqref{sec:means}. 
\begin{Corollary}
\label{propo:epf:props1}
Let $X \sim P_X$ be a random variable with equivalent probability function $\srprob[r]{P_X}$. Then: 
\begin{enumerate}

\item For all $r \in [-\infty, \infty]$, there holds that
	\begin{align}
		\min_k p_k= \srprob[-\infty]{P_X}  \leq \srprob[r]{P_X} \leq \max_k p_k= \srprob[\infty]{P_X} 
	\end{align}
	
%\item For all $p_k$ in $P_X$ there exists an $r \in [-\infty, \infty]$ such that $p_k = \srprob[r]{P_X}$\,. 
	
\item  If $P_X \equiv U_X$ the uniform over the same $\supp{P_X}$, then $\forall k, \forall r \in [-\infty, \infty], p_k = \srprob[r]{U_X} = \frac{1}{|\supp{P_X}|}$. 
	
\item if $P_X \equiv \delta_{X}^k$ the Kroneker delta centered on $x_k = X(\omega_k)$, then $\srprob{\delta_{X}^k} = u(r)$ where $u(r)$ is the step function. 
%$\srprob[\infty]{\delta_{X}^k} = 1$

\end{enumerate}
\end{Corollary}
\begin{proof}
Claims 1 and 2 issue directly from the properties of the entropies and the inverse to the logarithm. 
The last claims follows from Remark~\ref{rem:disc}.
%\todo[inline]{prove what happens to the other values of $r$. }
\end{proof}
And so, in their turn, the properties of R\'enyi entropy can be proven from those of the equivalent probability function and Hartley's generalized information function. %, being a dual isomorphim of positive semifields. 
%\begin{proof}
%To be done.
%\end{proof}

% % % % % % % % % % % % % % % % % % % % % % % % % % % % % % % % % % % % % % % % % % % % % % % % % % % % % % %
An interesting property might help recovering $P_X$ from the equivalent probability function:
\begin{Lemma}
\label{propo:epf:props2}
Let $X \sim P_X$ be a random variable with equivalent probability function $\srprob{P_X}$. Then: 
for every $p_k$ in $P_X$ there exists an $r_k \in [-\infty, \infty]$ such that $p_k = \srprob[r_k]{P_X}$\,. 
\end{Lemma}
\begin{proof}
This follows from the continuity of the means with respect to its parameters $\vec w$ and $\vec x$. 
\end{proof}

So if we could actually find those values $r_k, 1\leq k \leq n$ which return $p_k = \srprob[r]{P_X}$ we would be able to retrieve $P_X$ by sampling $\srprob[r_k]{P_X}$ in the appropriate values 
$P_X = \{ \srprob[r_k]{P_X}\}_{k=1}^n$. 
Since $n \geq 2$ we know that at least two of these values are $r = \pm \infty$ retrieving the value of the highest and lowest probabilities for $k=1$ and $k=n$ when they are sorted by increasing probability value. 

\begin{Example}[Continued]
Figure~\ref{fig:eq:prob}.\subref{fig:eq:prob:EPF} shows the equivalent probability function of the example in the previous section. 
The dual monotone behaviour with respect to that of the R\'enyi spectrum is clearly observable. 
We have also plotted over the axis at $r=0$ the original probabilities of the distribution to set it in the context of the properties in Corollary~\ref{propo:epf:props1} and Lemma~\ref{propo:epf:props2}.
\qed
\end{Example}

%\begin{proposition}
%To totally reconstruct a discrete probability distribution with $|\mathcal X| = n$ probabilities you need to know at least $n$ entropy values, of which at least two must the max and min entropies. 
%\end{proposition}

%% file: partitionPotential.tex
In the context of Information Theoretic Learning (ITL) the information potential is an important quantity~\cite[Ch.~2]{pri:10}. 
\begin{Definition}
Let $X \sim P_X$. %with R\'enyi spectrum $\srentropy[r]{P_X}$. 
Then the \emph{information potential} $\srpot{P_X}$ is 
\begin{align}
\label{eq:inf:pot}
\srpot{P_X} = E_{P_X}\{P_X^r\} = \sum_i \frac{p_i}{\sum_k p_k} p_i^r
\end{align}
\end{Definition}
Note that the original definition of the information potential was presented in terms of parameter $\alpha$ and for distributions with $\sum_k p_k = 1$ in which case $V_\alpha(P_X) = \srpot{P_X}$. 
Now, recall the conversion function in \eqref{eq:adapting} $\varphi'(h) = b^{-rh}$. The next lemma is immediate using it on \eqref{eq:sre:pot}.
\begin{Lemma}
\label{lemma:part:pot}
Let $X \sim P_X$. The \emph{information potential} is the $\varphi'$ image  of the shifted R\'enyi entropy 
\begin{align}
\srpot{P_X} = \varphi'(\srentropy{P_X})= b^{-r\srentropy{P_X}}
\end{align}
\end{Lemma}
\begin{proof}
%Since $\srentropy{P_X}$ is an entropy value, it may be directly transformed with $\varphi'$ to obtain the information potential. 
$\srpot{P_X} = b^{-r\srentropy{P_X}}=b^{\log_b (\sum_i \frac{p_i}{\sum_k p_k} p_i^r )} = \sum_i \frac{p_i}{\sum_k p_k} p_i^r = E_{P_X}\{P_X^r\}$
\end{proof}
\noindent 
Incidentally, \eqref{eq:pot:mean} gives the relation of the information potential and the generalized weighted means. 

%\todo{So far 21/09/18}
\begin{Remark}
The quantity in the RHS of \eqref{eq:inf:pot} is also the normalizing factor or \emph{partition function} of the moments of the distribution and, as such, appears explicitly  in the definition of  the escort probabilites~\eqref{eq:def:escort}. 
Usually other partition functions appear in the estimation of densities based in overt, e.g. maximum entropy~\cite{jay:96}, or in covert information criteria---e.g. Ising models~\cite{bec:sch:95}. 
\end{Remark}

%% file: discussionC4Shifting.tex
A number of decisions taken in the paper might seem arbitrary. In the following, we try to discuss these issues as well as alternatives left for future work. 

\subsubsection{Other reparameterization of the R\'enyi entropy}
Not only the parameter, but also de sign of the parameter is somewhat arbitray in the form of \eqref{eq:exponential}. 
%\begin{itemize}
%\item 
If we choose $r' = 1 - \alpha$ another generalization evolves that is, in a sense, symmetrical to the shifted R\'enyi entropy we have presented above, since $r' = -r$. This may be better or worse for the general formulae describing entropy, etc., but presents the problem that  it no longer aligns with Shannon's original choice of sign. The $r=0$ order R\'enyi entropy would actually be Boltzmann's, negative entropy or \emph{negentropy}~\cite{bri:62} and perhaps more suitable for applications in Thermodynamics~\cite{bec:sch:95}. 

%\item 
Yet another formulations suggest the use of $\alpha = 1/2$, equivalently $r=-1/2$ as the origin of the parameter~\cite{har:05}. 
%\end{itemize}
From our perspective, this suggests that the origin of the R\'enyi entropy can be chosen adequately in each application. 

\subsubsection{Other magnitudes around the R\'enyi entropy}

Sometimes the $p$-norm is used as a magnitude related to the R\'enyi entropy much as the information potential~\cite{pri:10} or directly seeing the relationship with the definition~\cite{bec:sch:95}.

\begin{Definition}
For a  set of non-negative numbers $\vec x = [x_i]_{i=1}^n \in [0,\infty)^n$ the $p$-norm, with $0 \leq p \leq \infty$ is
\begin{align}
|| \vec x ||_p = \left( \sum_i x_i^p\right)^\frac{1}{p}
\end{align}
\end{Definition}
\noindent 
A more general definition involves both positive and negative components for $\vec x$, as in normed real spaces, but this is not relevant to our purposes for non-negative measures. 

The $p$-norm has the evident problem that it is only defined for positive $p$ whereas \eqref{eq:adapting} proves than negative orders are meaningful and, indeed, interesting. 
A prior review of the negative orders can be found in~\cite{erv:har:14}. 

We believe this is yet one more advantage of the shifting of the R\'enyi index: that the relation with %the moments of the probability function, as captured by 
the equivalent probability function and the information potential---the moments of the distribution---are properly highlighted.

%\todo[inline]{Discuss later how this entropy changes the emphasis on the relation of the R\'en entropy from the p-norm  to the moments of the distribution, the moments being defined also for negative indices (whereas the norms are not).}

\subsubsection{Redundancy of the R\'enyi entropy}
\label{sec:Renyi:redundancy}

%\begin{Remark}
%\label{rem:Renyi:redundancy}
Lemma~\ref{lemma:ren:shan} proves that R\'enyi entropies are very redundant in the sense that given its value for a particular $r_0$ the rest can be written in terms of those entropies with different, but systematically related, $r$ index (see \S~\ref{sec:Renyi:redundancy}).
%\end{Remark}

In particular, equations~\eqref{eq:ren:shan:div} and~\eqref{eq:ren:shan:ent} in Lemma~\ref{lemma:ren:shan}, and~\eqref{eq:crosse:basis} in Lemma~\ref{lemma:crosse:basis} allow us to use a good estimator of Shannon's entropy to estimate the R\'enyi entropies and related magnitudes for all orders, special or not. 
Three interesting possibilities for this rewriting are:
\begin{itemize}
\item \emph{That everything can be written in terms of $r=0$, e.g. in terms of Shannon's entropy.}

This is made possible by the existence of estimators for Shannon's entropy and divergence. 

\item \emph{That everything can be written in terms of a finite $r \neq 0$, e.g. $r=1$.} 

This is possible by means of Properties~\ref{propo:prop:hmean}.\ref{prop:hmean:3} and~\ref{propo:prop:hmean}.\ref{prop:hmean:4} of the generalized power means. 
The work in~\cite{pri:10} is pointing this way
(perhaps including also $r=-1$, aka Hartley's) capitalizing on the fact that R\'enyi's entropy for data is well estimated for $r=1$, equivalently $\alpha=2$~\cite[\S~2.6]{pri:10}. 

%This is the approach  of~\cite{pri:10} where a specific estimate is given for each value of $r$ but is in practice used for $r=1$, equivalently $\alpha=2$. 

\item \emph{That everything can be written in terms of the extreme values of the entropy, e.g. $r = \pm \infty$.} 

This is suggested by Properties \ref{prop:prop:sE}.\ref{prop:prop:sE:p11} and~\ref{prop:prop:sE}.\ref{prop:prop:sE:p12}. Supposing we had a way to estimate either $\srentropy[-\infty]{P_X}$ or $\srentropy[\infty]{P_X}$. Then by a divide-and-conquer type of approach it would be feasible to extract all the probabilities of a distribution out of its R\'enyi entropy function. 

\end{itemize}

\subsubsection{The algebra of entropies}
\label{sec:semifields}
Technically, the completed non-negative reals $\nnR$, where the means are defined, carry a complete positive semifield structure~\cite{gon:min:08}. This is an algebra similar to a real-valued field but the inverse operation to addition, e.g. subtraction, is missing. 

There are some technicalities involving writing the results of the operations of the extremes of the semifields---e.g. multiplication of $0$ and $\infty$---and this makes writing closed expressions for the means with extreme values of $\vec w$ or $\vec x$ complicated.
 A sample of this is the plethora of conditions on Property~\ref{propo:prop:hmean}.\ref{prop:hmean:5}. 
 An extended notation, pioneered by Moreau~\cite{mor:70}, is however capable of writing a closed expression for the means~\cite{val:pel:17c}. 

Furthermore, taking (minus) logarithms and raising to a real power are isomorphism of semifields, so that the R\'enyi entropies inhabit a different positive semifield structure~\cite{val:pel:17c}.  The graph of these isomorphic structures can be seen in Figure~\ref{fig:circaREntropy}.\subref{fig:circaREntropy:domains}. 
This means that some of the intuitions about operating with entropies are misguided. We believe that failing to give a meaning to the R\'enyi entropies with negative orders might have been caused by this. 

\subsubsection{Shifted R\'enyi entropies on continuous distributions}

%\todo[inline]{FVA 6.11.18: Include a subsection on the shifted continuous entropy?}

%\todo[inline]{Introduce the concept o the continuous means already described by de Finetti~\cite{fin:31}. Development in~\cite{har:lit:pol:52}.}

The treatment we use here may be repeated on continuous measures, but the definitions of Shannon~\cite{kol:56,cov:tho:06} and R\'enyi~\cite{ren:70} entropies in such case run into technical difficulties solved, typically, by a process of discretization~\cite{jiz:ari:04}. 

Actually we believe that the shifting would also help in this process: a form for the generalised weighted continuous means was long ago established~\cite{fin:31} and technically solved by a change of concept and Lebesgue-Stieltjes integration instead of summation~\cite[Ch.~VI]{har:lit:pol:52}. 

Our preliminary analyses show that the relationship with the means given by \eqref{eq:sRenyE} also holds, and this would mean that the shifting---in aligning the R\'enyi entropies with the (generalize weighted) continuous means---leverages the theoretical support of the latter to sustain the former.  

\begin{Definition}[Continuous weighted $f$-mean]
Let $\Phi(\xi)$ be a measure and let $f$ be a monotonic function of $\xi$ with inverse $f^{-1}$. Then a continuous version of~\eqref{eq:KN:formula} is:
\begin{align*}
\hmean[f]{\Phi}{\xi} &= f^{-1} \left\{\int f(\xi) d\Phi(\xi)\right\}
\end{align*}
understood as a Lebesgue-Stieltjes integral. 
\end{Definition}
This definition was already proposed by De Finetti~\cite{fin:31} based upon the works of Bonferroni and Kolmogorov and thoroughly developed in~\cite[Ch.~VI]{har:lit:pol:52} in connection to the discrete means. 
With $f(x) = x^r$ the continuous H\"older means $\hmean[r]{\Phi}{\xi}$ appear.
Furthermore De Finetti found~\cite[\#8]{fin:31} that the form of the $f$ continuous, monotone function $f$ must be
\begin{align*}
f(x) &= a \int \gamma(x) dx + b &
&\text{for arbitrary }a,b (a \neq 0)
\end{align*}
similar to what R\'enyi found later for the Shannon entropy.

It is easy to see that an analogue definition of the shifted R\'enyi entropy but for a continuous probability density $p_X$ with $dp_X(x) =  p_X(x) dx$~\cite{bec:sch:95,jiz:ari:04} is 
\begin{align}
\label{eq:sRenyi:cont}
\tilde h(p_X) = \frac{-1}{r} \log \int p_X(x) p_X^r(x) dx = - \log \hmean[r]{p_X}{p_X}
\end{align}
again with the distribution acting as weight and averaged quantity. 
Compare this to one of the standard forms of the \emph{differential R\'enyi entropy}~\cite{erv:har:14}:
\begin{align*}
h(p_X) = \frac{1}{1-\alpha} \ln \int p_X^\alpha(x) dx
\end{align*}
The investigation of the properties of \eqref{eq:sRenyi:cont} is left pending for future work, though. 

\subsubsection{Pervasiveness of R\'enyi entropies}
% As a measure of complexity
%
Apart form the evident applications to 
signal processing and communications~\cite{pri:10}, 
physics~\cite{bec:sch:95} and 
cognition~\cite{say:18},  %Machine Intelligence mentioned above, 
the R\'enyi entropy is a measure of diversity in several disciplines~\cite{zha:gra:16}. %(Economics, Ecology, etc.)
It is not unconceivable that its if applicability comes from the same properties stemming from the means that we have expounded in this paper as applied to positive distributions e.g. of wealth in a population, or energy in a community, then the expression to be used is~\eqref{eq:mass:edispl}. 

%\subsubsection{General discussion}
%
%\todo[inline]{Issue with the escort probabilities: they are obtained prior to any consideration of the R\'enyi entropies when considering the derivatives of generalized means. }

%\todo[inline]{Relations to the semifield and forward references to EntropiesInSemfields?.}

%% file: conclusionsC4S.tex
%\subsubsection{The case for shifting the R\'enyi entropy}
In this paper we have advocated for the shifting of the traditional R\'enyi entropy order from a parameter $\alpha$ to $r = \alpha - 1$. 
The shifting of the R\'enyi entropy and divergence is motivated by a number of results:
%Hence, the first reason in support of the shifting is:
\begin{itemize}
\item  It aligns them with the power means and explains the apparition of the escort probabilities. Note that the importance of the escort probabilities is justified independently of their link to the means in the shifted version of entropy~\cite{bec:sch:95}. 

\item It highlights the Shannon entropy $r=0$ in the role of the ``origin"   of entropy orders, just as the geometric means is a particular case of the weighted averaged means. This consideration is enhanced by the existence of a formula allowing us to  rewrite every other order as a combination of Shannon entropies and cross entropies of escort probabilities of the distribution. 

\item The shifting of the R\'enyi entropy aligns it with the moments of the distribution, thus enabling new insights into the moments' problem. 

\item It makes the relation between the divergence and the entropy more ``symmetrical''. 

\item It highlights the ``information spectrum'' quality of the R\'enyi entropy measure for fixed $P_X$.

%\item It explains entropies as an application in semifield algebra, perhaps opening new insights and avenues of research. 
\end{itemize}

The shifting might or might not be justified by applications. 
If the concept of the means is relevant in the application, 
%If manipulation of the formulae are needed to highlight the relation to the means, 
we recommend the shifted formulation. 

%% file: KNmean.tex
% % % % % % % % % % % % % % % % % % % % % % % % % % % % % % % % % % % %
% Most of the material in this section comes from~\cite{har:lit:pol:52}. 
% % % % % % % % % % % % % % % % % % % % % % % % % % % % % % % % % % % %
%
The following is well-known since~\cite{fin:31,har:lit:pol:52}.  
\begin{Definition}
\label{def:KN}
Given an invertible real function $f: \mathbb R \rightarrow \mathbb R$ the \emph{Kolmogorov-Nagumo mean of a set of non-negative numbers} $\vec x = [x_i]_{i=1}^n \in [0,\infty)^n$ is
\begin{align}
\label{eq:KN}
KN_f(\vec x) = f^{-1}(\sum_{i=1}^n \frac{1}{n} f(x_i))\,.
\end{align}
\end{Definition}
%From now on, 
%let $\vec w=[w_i]_{i=1}^n \in (\nnR)^n$ and $\vec x=[x_i]_{i=1}^n\in (\nnR)^n$ be equally long, co-indexed sequences of non-negative numbers. 
Definition~\eqref{def:KN} is an instance of the following formula to work out the \emph{weighted $f$-mean} with a set of finite, non-negative weights, $\vec w \in [0,\infty)$
%$w = [w_i]_{i=1}^n \in (\nonnegativeR)^n$ 
\begin{align}
\label{eq:KN:formula}
M_f(\vec w,\vec x) = f^{-1}(\sum_{i=1}^n \frac{w_i}{\sum_k w_k} f(x_i))\,.
\end{align}

Our interest in~\eqref{eq:KN:formula} lies in the fact that Shannon's and Renyi's entropies can be seen as special cases of it, which makes its properties specially interesting.  
\begin{Proposition}[Properties of the Kolmogorov-Nagumo means]
\label{propo:KN}
%Let $\vec w \in (0,\infty)^n, \vec x \in [0,\infty)^n$ and $r,s \in \mathbb R\setminus0$\,.  
Let $\vec x, \vec w \in [0,\infty)^n$. 
The following conditions hold if and only if there is a strictly monotonic and continuous function $f$ such that~\eqref{eq:KN} holds.

\begin{enumerate}
\item \label{prop:knmean:9} \emph{Continuity and strict monotonicity in all coordinates.}

\item
\label{prop:knmean:1}
\emph{(Symmetry or permutation invariance)}  
Let $\sigma$ be a permutation, then 
$\hmean[f]{\vec w}{\vec x} = \hmean[f]{\sigma(\vec w)}{\sigma(\vec x)}$\,.

\item \label{prop:knmean:8} 
\emph{(Reflexivity)} 
The mean of a series of constants is the constant itself:
\[
\hmean[f]{\vec w}{\{k\}_{i=1}^n} = k
\]

\item \label{prop:knmean:5}
\emph{(Blocking)} 
	The computation of the mean can be split into computations of equal size sub-blocks.
	
\item \label{prop:knmean:4}
	\emph{(Associativity)} Replacing a $k$-subset of the $x$ with their partial mean in the same  multiplicity does not change the overall mean. 
\end{enumerate}
\end{Proposition}
For a minimal axiomatization, Blocking and Associativity are redundant. 
A review of the axiomatization of these and other properties can be found in~\cite{mul:par:93}.

%% file: postulatesShannon.tex
It is important to recall that Shannon set out to define the \emph{amount of} information, discarding any notions of \emph{information} itself. Both notions should be distinguished clearly for methodological reasons, but can be ignored for applications that deal only with quantifying information. %, like Information Theory itself. 

%For that purpose, recall 
% Faddev the father, the mathematician, not the son, the physicist.
Recall the Faddeev postulates for the generalization of Shannon's entropy~\citep[Chap. IX. \S2]{ren:70}:
%\emph{ %
\begin{enumerate} %[Postulates of Information]
\item The \emph{amount of information} $\sentropy{P}$ of a sequence $P=[p_k]_{k=1}^{n}$ of $n$ numbers 
%where the notation means $\{p_k\}_{k=1}^{n} = \{p_k \mid 1 \leq k  \leq n\}$, and it 
is a symmetric function of this set of values $\sentropy{P} = \sentropy{\sigma(P)} = \sentropy{\{p_k\}_{k=1}^{n}}$, where $\sigma$ is any permutation of $n$-elements. 

\item $\sentropy{\{p, 1-p\}}$ is a continuous function of $p, 0 \leq p \leq 1$\,.

\item $\sentropy{\{\frac{1}{2}, \frac{1}{2}\}} = 1$\,.

\item The following relation holds:
\begin{align}
\sentropy{\{p_1, p_2, \ldots, p_n\}} = \sentropy{\{p_1 + p_2, \ldots, p_n \}} + (p_1 + p_2) \sentropy{\{\frac{p_1}{p_1 + p_2}, \frac{p_2}{p_1 + p_2}\}}
\end{align}
\end{enumerate}
%} %Emphasis

These postulates lead to Shannon's entropy for $X \sim P_X$ with binary logarithm~\cite{ren:70}
\begin{align}
\label{eq:Sentropy}
\sentropy{P_X}= E_{P_X}\{-\log P_X\} = - \sum_k p_k \log p_k
\end{align}

%% file: case4shifting-Arxiv.bbl
\begin{thebibliography}{-------}
\providecommand{\natexlab}[1]{#1}

\bibitem[R\'enyi(1961)]{ren:61}
R\'enyi, A.
\newblock On measures of entropy and information.
\newblock  Fourth Berkeley Symposium,  1961, pp. 547--561.

\bibitem[Shannon and Weaver(1949)]{sha:wea:49}
Shannon, C.; Weaver, W.
\newblock {\em {A mathematical model of communication}}; The University of
  Illinois Press,  1949.

\bibitem[Shannon(1948{\natexlab{a}})]{sha:48a}
Shannon, C.E.
\newblock {A mathematical theory of Communication}.
\newblock {\em The Bell System Technical Journal} {\bf 1948}, {\em
  XXVII},~379--423.

\bibitem[Shannon(1948{\natexlab{b}})]{sha:48b}
Shannon, C.E.
\newblock {A mathematical theory of communication}.
\newblock {\em The Bell System Technical Journal} {\bf 1948}, {\em
  XXVII},~623--656.

\bibitem[Van~Erven and Harremo{\"e}s(2014)]{erv:har:14}
Van~Erven, T.; Harremo{\"e}s, P.
\newblock {R{\'e}nyi divergence and Kullback-Leibler divergence}.
\newblock {\em IEEEE Transaction on Information Theory} {\bf 2014}.

\bibitem[Hardy \em{et~al.}(1952)Hardy, Littlewood, and
  P{\'o}lya]{har:lit:pol:52}
Hardy, G.H.; Littlewood, J.E.; P{\'o}lya, G.
\newblock {\em {Inequalities}}; Cambridge University Press,  1952.

\bibitem[Kitagawa(1934)]{kit:34}
Kitagawa, T.
\newblock On Some Class of Weighted Means.
\newblock {\em Proceedings of the Physico-Mathematical Society of Japan. 3rd
  Series} {\bf 1934}, {\em 16},~117--126.
\newblock
  doi:{\changeurlcolor{black}\href{https://doi.org/10.11429/ppmsj1919.16.0_117}{\detokenize{10.11429/ppmsj1919.16.0_117}}}.

\bibitem[Beck and Sch{\"o}gl(1995)]{bec:sch:95}
Beck, C.; Sch{\"o}gl, F.
\newblock {\em {Thermodynamics of Chaotic Systems: An Introduction}}; Cambridge
  University Press,  1995.

\bibitem[Renyi(1970)]{ren:70}
Renyi, A.
\newblock {\em {Probability Theory}}; Courier Dover Publications,  1970.

\bibitem[Jizba and Arimitsu(2004)]{jiz:ari:04}
Jizba, P.; Arimitsu, T.
\newblock {The world according to R{\'e}nyi: thermodynamics of multifractal
  systems}.
\newblock {\em Annals of Physics} {\bf 2004}, {\em 312},~17--59.

\bibitem[Bickel \em{et~al.}(1975)Bickel, Hammel, and O'Connell]{bic:ham:oco:75}
Bickel, P.J.; Hammel, E.A.; O'Connell, J.W.
\newblock Sex bias in graduate admissions: Data from Berkeley.
\newblock {\em Science} {\bf 1975}, {\em 187},~398--403.

\bibitem[Principe(2010)]{pri:10}
Principe, J.C.
\newblock {\em {Information Theoretic Learning}}; Information Science and
  Statistics, Springer: New York,  2010.

\bibitem[Jaynes(1996)]{jay:96}
Jaynes, E.T.
\newblock {\em {Probability theory: The logic of science}}; Cambridge
  University Press,  1996.

\bibitem[Brillouin(1962)]{bri:62}
Brillouin, L.
\newblock {\em {Science and information theory}}; Second edition, Academic
  Press, Inc., Publishers, New York,  1962.

\bibitem[Harremo{\"e}s(2005)]{har:05}
Harremo{\"e}s, P.
\newblock {Interpretations of R{\'e}nyi entropies and divergences}.
\newblock {\em Physica A: Statistical Mechanics and its Applications} {\bf
  2005}, {\em 365},~57--62.

\bibitem[Gondran and Minoux(2008)]{gon:min:08}
Gondran, M.; Minoux, M.
\newblock {\em Graphs, Dioids and Semirings. New Models and Algorithms.};
  Operations Research Computer Science Interfaces series., Springer,  2008.

\bibitem[Moreau(1970)]{mor:70}
Moreau, J.J.
\newblock {Inf-convolution, sous-additivit{\'e}, convexit{\'e} des fonctions
  num{\'e}riques (in French)}.
\newblock {\em J Math Pures Appl(9)} {\bf 1970}.

\bibitem[Valverde~Albacete and Pel\'aez-Moreno(2017)]{val:pel:17c}
Valverde~Albacete, F.J.; Pel\'aez-Moreno, C.
\newblock Entropy operates in Non-Linear Semifields.
\newblock {\em Arxiv} {\bf 2017}.

\bibitem[Kolmogorov(1956)]{kol:56}
Kolmogorov, A.
\newblock {On the Shannon theory of information transmission in the case of
  continuous signals}.
\newblock {\em IRE Transactions on Information Theory} {\bf 1956}, {\em
  2},~102--108.

\bibitem[Cover and Thomas(2006)]{cov:tho:06}
Cover, T.M.; Thomas, J.A.
\newblock {\em {Elements of Information Theory }}, 2nd ed.; {John Wiley \&
  Sons},  2006.

\bibitem[De~Finetti(1931)]{fin:31}
De~Finetti, B.
\newblock {Sul concetto di media}.
\newblock {\em Giornale dellIstituto Italiano degli Attuari} {\bf 1931}, {\em
  II},~369--396.

\bibitem[Sayood(2018)]{say:18}
Sayood, K.
\newblock {Information Theory and Cognition: A Review}.
\newblock {\em Entropy} {\bf 2018}, {\em 20},~1--19.

\bibitem[Zhang and Grabchak(2016)]{zha:gra:16}
Zhang, Z.; Grabchak, M.
\newblock Entropic representation and estimation of diversity indices.
\newblock {\em Journal of Nonparametric Statistics} {\bf 2016}, {\em
  28},~563--575.

\bibitem[Muliere and Parmigiani(1993)]{mul:par:93}
Muliere, P.; Parmigiani, G.
\newblock {Utility and means in the 1930s}.
\newblock {\em Statistical Science} {\bf 1993}.

\end{thebibliography}
